\documentclass[preprintnumbers,amsmath,amssymb,twocolumn]{revtex4-1}

\usepackage{graphicx}
\usepackage{dcolumn}
\usepackage{bm}
\usepackage{tikz}
\usepackage{pgfplots}
\usepackage{latexsym,epsfig,color,url,bbm,psfrag}
\usepackage[caption=false]{subfig}
\usepackage{chngcntr}
\usepackage{booktabs}
\usepackage[english]{babel}
\usepackage{hyperref}
\usepackage{todonotes}
\usepackage{amsmath,amsthm}
\usepackage{xr}
\usepackage{xcite}
\newcommand*{\swap}[2]{#2#1}

\newcommand{\prob}{\mathbb{P}}
\newcommand{\Prob}[1]{\prob\left(#1\right)}

\newcommand{\me}{{\rm e}}
\newcommand{\w}{h}
\newcommand{\dd}{{\rm d}}
\newcommand{\mean}[1]{\langle #1 \rangle}
\newcommand{\hs}{h_s}
\newcommand{\hst}{h_s^2}

\newtheorem{proposition}{Proposition}

\newtheorem{lemma}{Lemma}

\allowdisplaybreaks
\begin{document}

\title{Clustering spectrum of scale-free networks}

 \author{Clara Stegehuis}

 \author{Remco van der Hofstad}

\author{A.J.E.M. Janssen}
\author{Johan S.H. van Leeuwaarden}%

\affiliation{Eindhoven University of Technology, Department of Mathematics and Computer Science, P.O. Box 513, 5600 MB Eindhoven, The Netherlands}

\date{\today}

             \graphicspath{{Figures/}}

\begin{abstract}
Real-world networks often have power-law degrees and scale-free properties such as ultra-small distances and
ultra-fast information spreading. In this paper, we study a third universal property: three-point correlations that suppress the creation of triangles and signal the presence of hierarchy. We quantify this property in terms of $\bar c(k)$, the probability that two neighbors of a degree-$k$ node are neighbors themselves. We investigate how the clustering spectrum $k\mapsto\bar c(k)$ scales with $k$ in the hidden variable model and show that $c(k)$ follows a {\it universal curve} that consists of three $k$-ranges where $\bar c(k)$ remains flat, starts declining, and eventually settles on a power law $\bar c(k)\sim k^{-\alpha}$ with $\alpha$ depending on the power law of the degree distribution. 
We test these results against ten contemporary real-world networks and explain analytically why the universal curve properties only reveal themselves in large networks.
\end{abstract}

\pacs{89.75.-k Complex systems, 64.60.aq Networks}
\maketitle

\section{Introduction}
 Most real-world networks have power-law degrees, so that the proportion of nodes having $k$ neighbors scales as $k^{-\tau}$ with exponent $\tau$ between 2 and 3 \cite{albert1999,faloutsos1999,jeong2000,vazquez2002}. Power-law degrees imply various intriguing scale-free network properties, such as ultra-small distances~\cite{hofstad2007, newman2001} and the absence of percolation thresholds when $\tau<3$~\cite{janson2009b, pastor2001}. Empirical evidence has been matched by random graph null models that are able to explain mathematically why and how these properties arise. This paper deals with another fundamental property observed in many scale-free networks related to three-point correlations that suppress the creation of triangles and signal the presence of hierarchy. We quantify this property in terms of the {\it clustering spectrum}, the function $k\mapsto \bar c(k)$ with $\bar c(k)$ the probability that two neighbors of a degree-$k$ node are neighbors themselves.

\begin{figure}[h]
 \centering
		\includegraphics[width=0.8\linewidth]{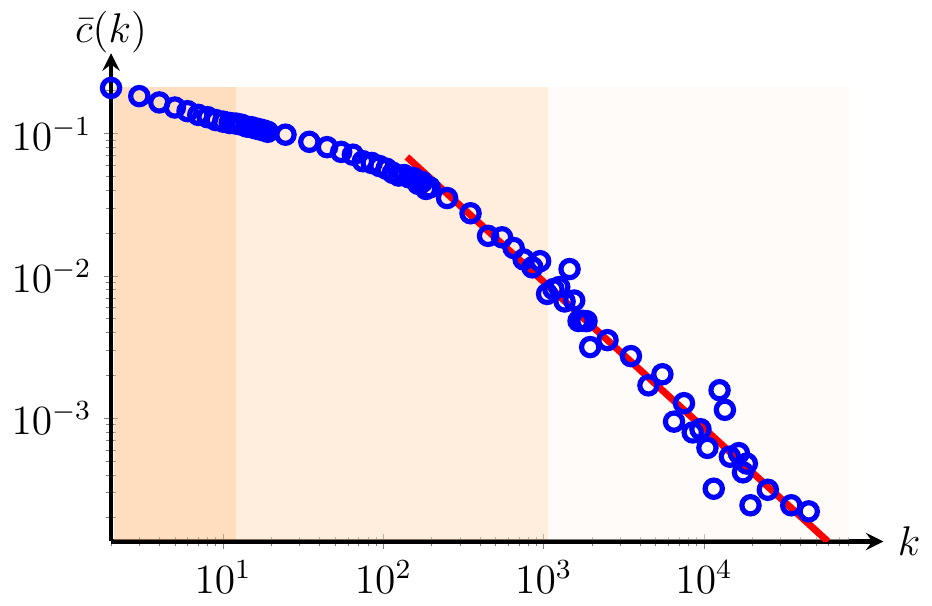}
\caption{$\bar c(k)$ for the YouTube social network}
		\label{fig:chas}
\end{figure}

 In {\it uncorrelated} networks the clustering spectrum $\bar c (k)$ remains constant and independent of $k$. However, the majority of real-world networks have spectra that decay in $k$, as first observed in technological networks including the Internet~\cite{pastor2001b,ravasz2003}. Figure~\ref{fig:chas} shows the same phenomenon for a social network: YouTube users as vertices, and edges indicating friendships between them~\cite{snap}.

\begin{figure*}
 	\subfloat[]{%
 		\centering
 		\includegraphics[width=0.33\linewidth]{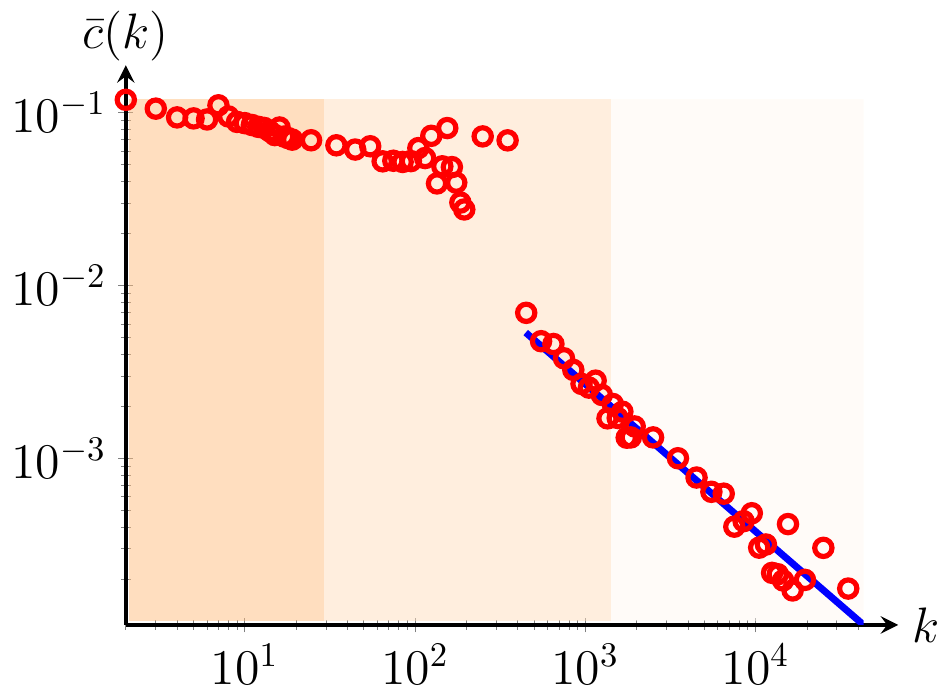}
 		\label{fig:chhudong}
 	}%
 	\subfloat[]{%
 		\centering
 		\includegraphics[width=0.33\linewidth]{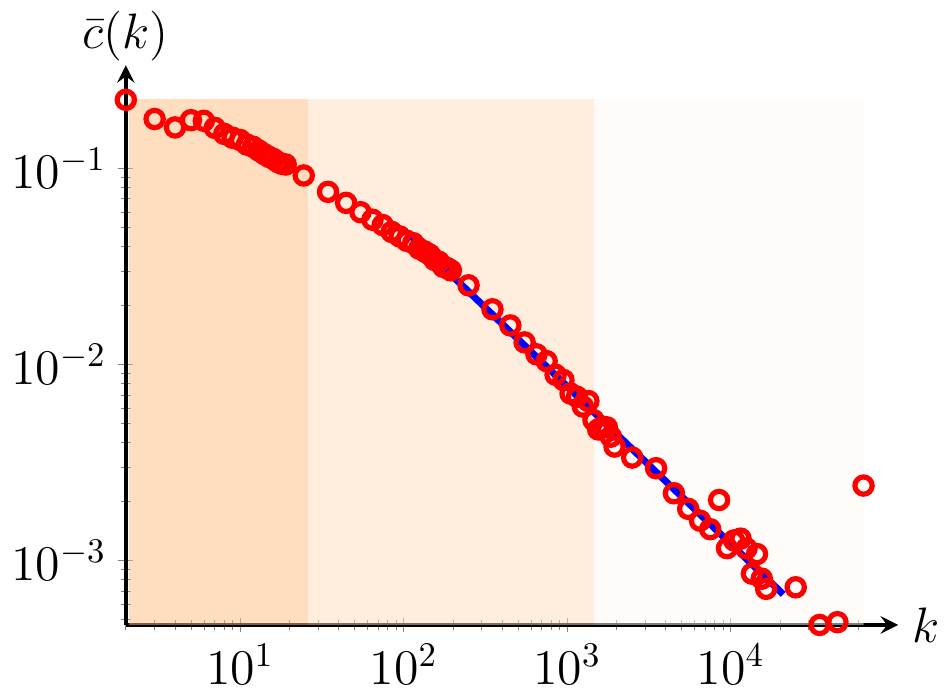}
 		\label{fig:chbaidu}
 	}
 	\subfloat[]{
 		\centering
 		\includegraphics[width=0.33\linewidth]{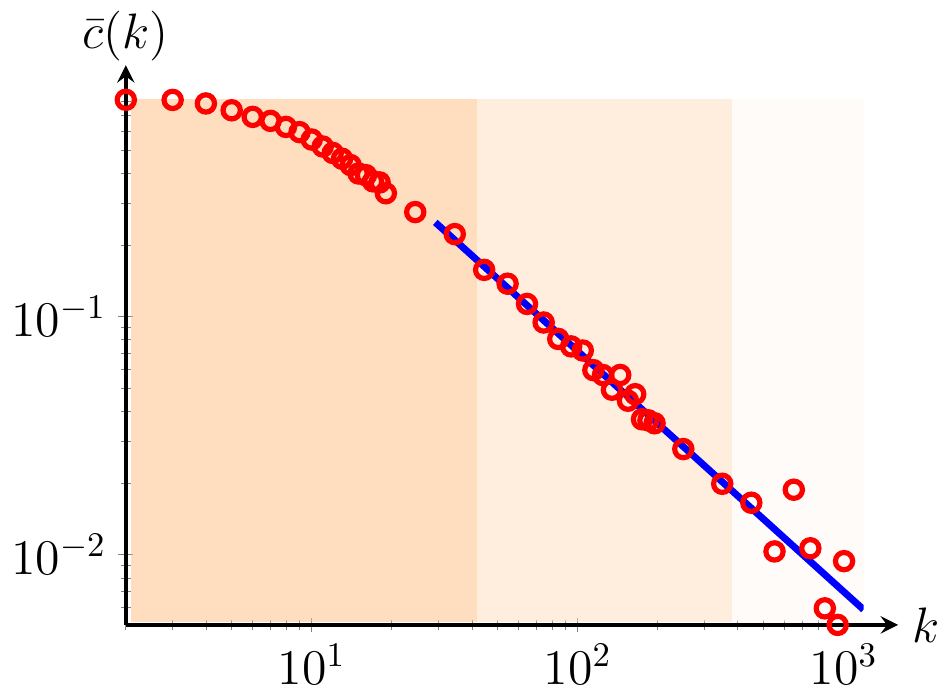}
 		\label{fig:chwordnet}
 	}

 	\subfloat[]{
 		\centering
 		\includegraphics[width=0.33\linewidth]{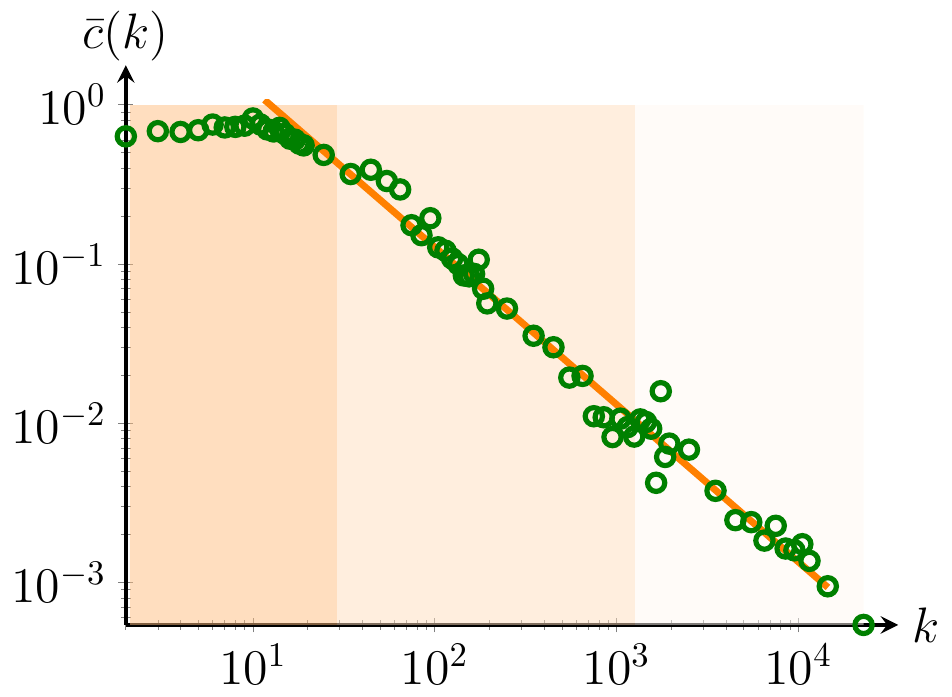}
 		\label{fig:chtrec}
	}
 	\subfloat[]{
 		\centering
 		\includegraphics[width=0.33\linewidth]{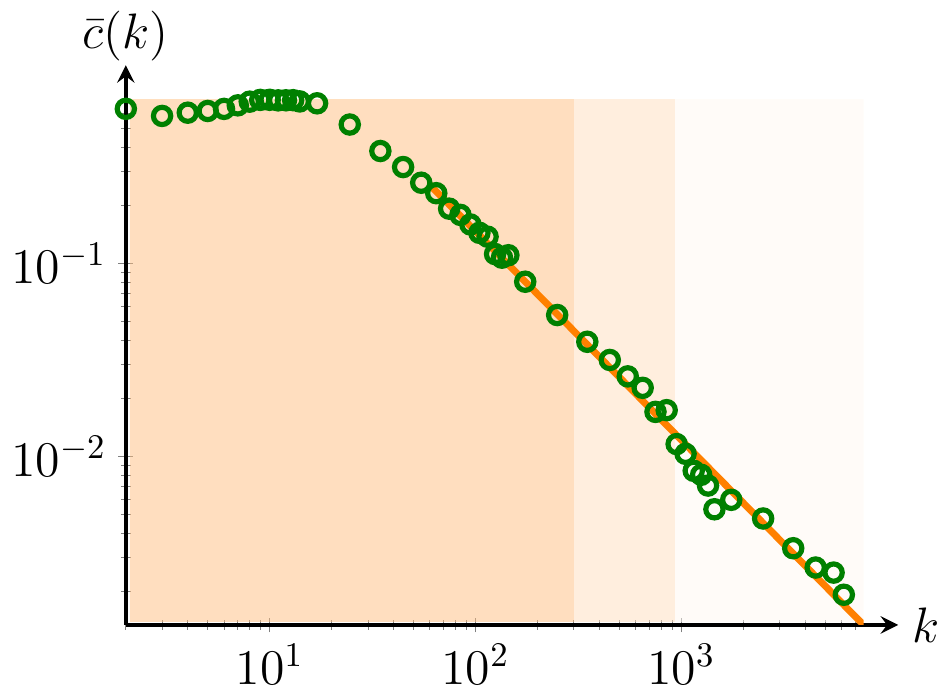}
 		\label{fig:chgoogle}
 	}
 	\subfloat[]{
 		\centering
 		\includegraphics[width=0.33\linewidth]{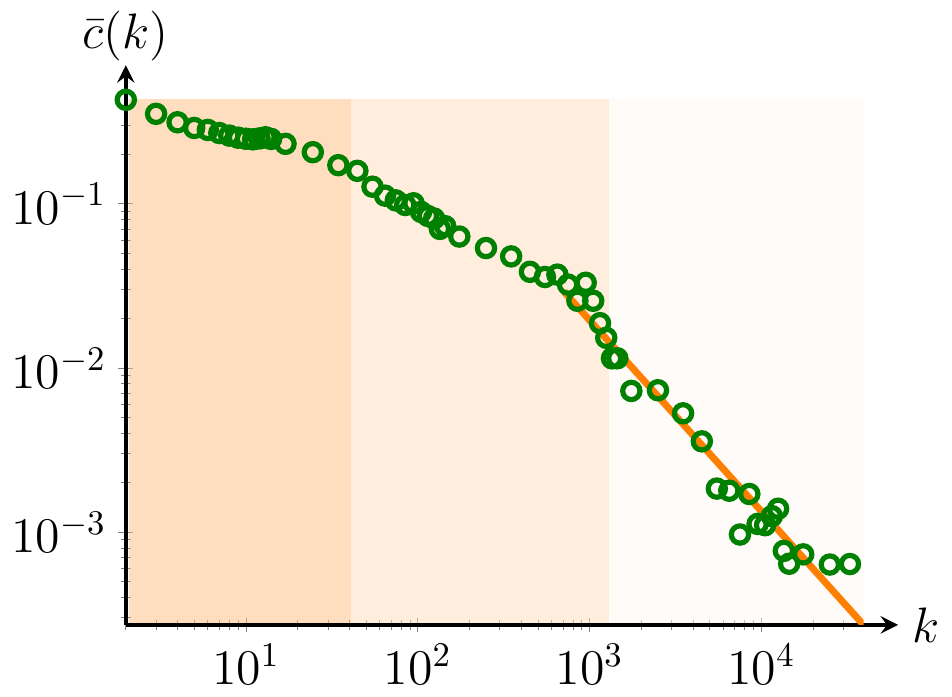}
 		\label{fig:chwiki}
 	}
 	
 	\subfloat[]{%
 		\centering
 		\includegraphics[width=0.33\linewidth]{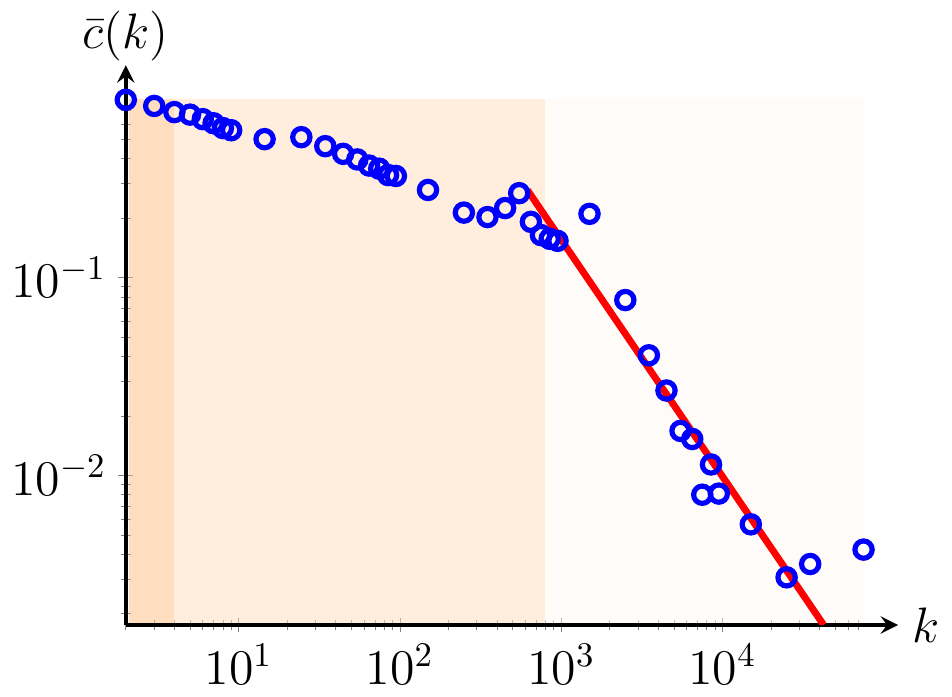}
 		\label{fig:chcatster}
 	}%
 	\subfloat[]{
 		\centering
 		\includegraphics[width=0.33\linewidth]{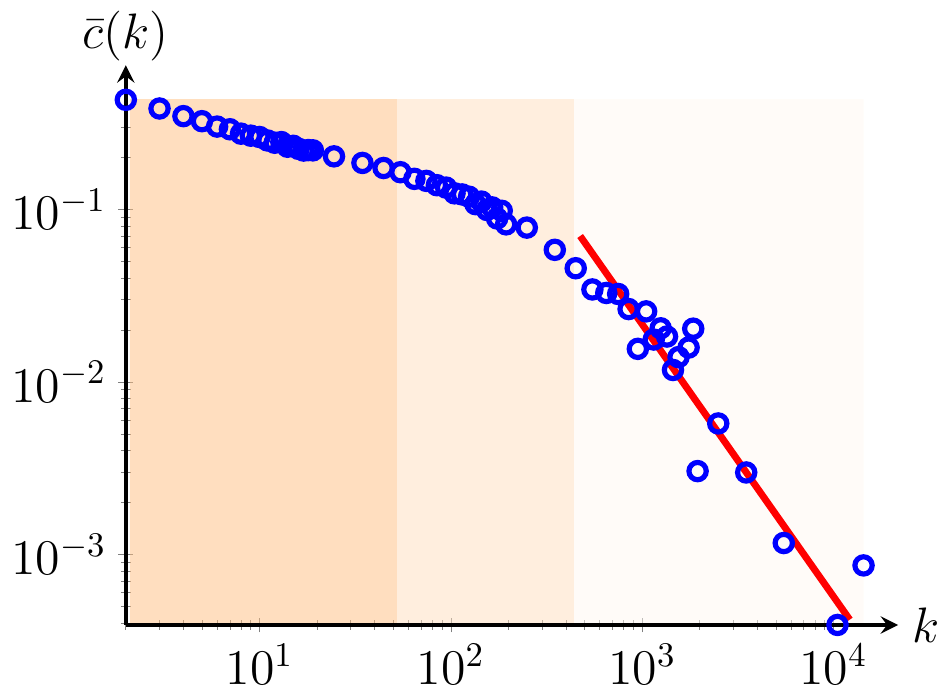}
 		\label{fig:chgowalla}
 	}
 	\subfloat[]{
 		\centering
 		\includegraphics[width=0.33\linewidth]{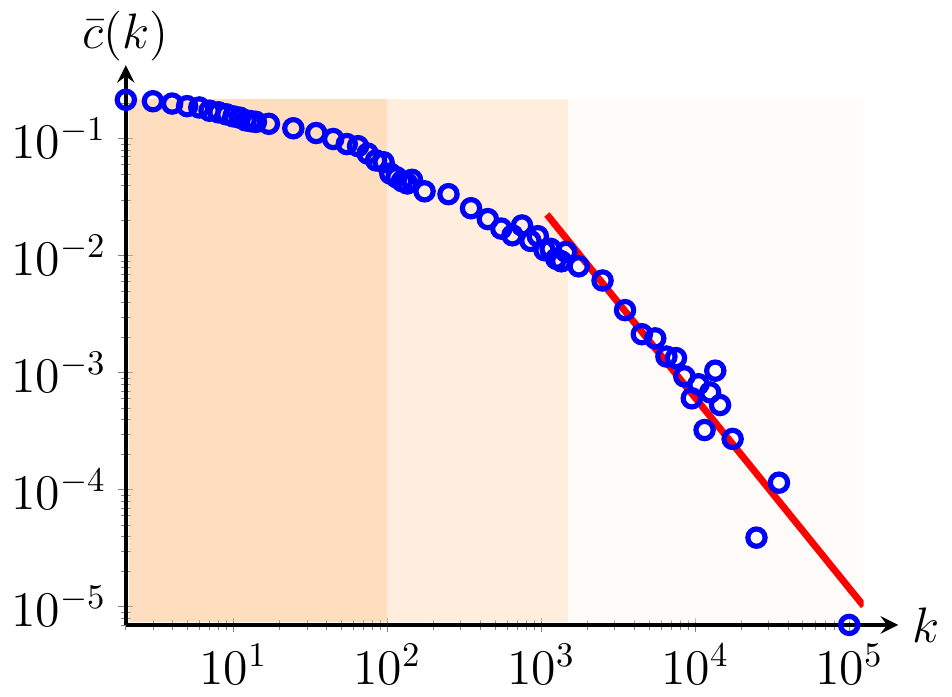}
 		\label{fig:chyou}
 	}
 	\caption{$\bar c(k)$ for several information (red), technological (green) and social (blue) real-world networks. (a) Hudong encyclopedia~\cite{niu2011}, (b) Baidu encyclopedia~\cite{niu2011}, (c) WordNet~\cite{miller1998}, (d) TREC-WT10g web graph~\cite{bailey2003}, (e) Google web graph~\cite{snap}, (f) Internet on the Autonomous Systems level~\cite{snap}, (g) Catster/Dogster social networks~\cite{konect}, (h) Gowalla social network~\cite{snap}, (i) Wikipedia communication network~\cite{snap}. The different shadings indicate the theoretical boundaries of the regimes as in Fig.~\ref{fig:curve}, with $N$ and $\tau$ as in Table~\ref{tab:data}.
 	}
 	\label{fig:ch}
 \end{figure*}

 Close inspection suggests the following properties, not only in Fig.~\ref{fig:chas}, but also in the nine further networks in Fig.~\ref{fig:ch}. The right end of the spectrum appears to be of the power-law form $k^{-\alpha}$; approximate values of $\alpha$ give rise to the dashed lines; (ii) The power law is only approximate and kicks in for rather large values of $k$. In fact, the slope of $\bar c (k)$ decreases with $k$; (iii) There exists a transition point: the minimal degree as of which the slope starts to decline faster and settles on its limiting (large $k$) value. 
 
For scale-free networks a decaying $\bar c(k)$ is taken as an indicator for the presence of modularity and hierarchy~\cite{ravasz2003}, architectures that can be viewed as collections of subgraphs with dense connections within themselves and  sparser ones between them. The existence of clusters of dense interaction signals hierarchical or nearly decomposable structures. When the function $\bar c(k)$ falls off with $k$,  low-degree vertices have relatively high clustering coefficients, hence creating small modules that are connected through triangles. In contrast, high-degree vertices have very low clustering coefficients, and therefore act as bridges between the different local modules. This also explains why $\bar c(k)$ is not just a local property, and when viewed as a function of $k$, measures crucial mesoscopic network properties such as modularity, clusters and communities. The behavior of $\bar c(k)$ also turns out to be a good predictor for the macroscopic behavior of the network. Randomizing real-world networks while preserving the shape of the $\bar c(k)$ curve produces networks with very similar component sizes as well as similar hierarchical structures as the original network~\cite{colomer2013}. Furthermore, the shape of $\bar c(k)$ strongly influences the behavior of networks under percolation~\cite{serrano2006}. This places the $\bar c(k)$-curve among the most relevant indicators for structural correlations in network infrastructures.

In this paper, we obtain a precise characterization of clustering in the hidden variable model, a tractable random graph null model. We start from an explicit form of the $\bar c(k)$ curve for the hidden variable model~\cite{boguna2003,serrano2007,dorogovtsev2004}. 
We obtain a detailed description of the $\bar c(k)$-curve in the large-network limit that provides rigorous underpinning of the empirical observations (i)-(iii). We find that the decay rate in the hidden variable model is significantly different from the exponent $\bar c(k)\sim k^{-1}$ that has been found in a hierarchical graph model~\cite{ravasz2003} as well as in the preferential attachment model~\cite{krot2015}  and a preferential attachment model with enhanced clustering~\cite{szabo2003}. Furthermore, we show that before the power-law decay of $\bar c(k)$ kicks in, $\bar c(k)$ first has a constant regime for small $k$, and a logarithmic decay phase. This characterizes the entire clustering spectrum of the hidden variable model.

This paper is structured as follows. Section \ref{sec:hidden} introduces the random graph model and its local clustering coefficient. Section \ref{sec:three} presents the main results for the clustering spectrum. Section \ref{sec:deep} explains the shape of the clustering spectrum in terms of an energy minimization argument, and Section \ref{sec:deep2} quantifies how fast the limiting clustering spectrum arises as function of the network size. We conclude with a discussion in Section  \ref{sec:disc} and present all mathematical derivations of the main results in the appendix. 

\section {Hidden variables}\label{sec:hidden}
As null model we employ the hidden variable model~ \cite{boguna2003,park2004,bollobas2007,britton2006,norros2006}.
Given $N$ nodes, hidden variable models are defined as follows.  Associate to each node a hidden variable $h$ drawn from a given probability distribution function 
\begin{equation}\label{eq:rhoh}
\rho(h)=Ch^{-\tau}
\end{equation}
for some constant $C$. Next join each pair of vertices independently according to a given probability $p(h,h')$ with $h$ and $h'$ the hidden variables associated to the two nodes. 
Many networks can be embedded in this hidden-variable framework, but particular attention goes to the case in which the hidden variables have themselves the structure of the degrees of a real-world network. In that case the hidden-variable  model puts soft constraints on the degrees, which is typically easier to analyze than hard constraints as in the configuration model~\cite{clauset2009,newman2003book,vazquez2002,dhara2016}.
Chung and Lu \cite{chung2002} introduced the hidden variable model in the form
\begin{equation}\label{c1}
p(h,h')\sim \frac{h h'}{N \mean{h}},
\end{equation}
so that the expected degree of a node equals its hidden variable.

We now discuss the structural and natural cutoff, because both will play a crucial role in the description of the clustering spectrum. The structural cutoff is defined as the largest possible upper bound on the degrees required to guarantee single edges, while the natural cutoff characterizes the maximal degree in a sample of $N$ vertices. For scale-free networks with exponent $\tau\in(2,3]$ the structural cutoff scales as $\sqrt{N}$ while the natural cutoff scales as $N^{1/(\tau-1)}$, which gives rise to structural negative correlations and possibly other finite-size effects. If one wants to avoid such effects, then the maximal value of the product $h h'$ should never exceed $N \mean{h}$, which can be guaranteed by the assumption that the hidden degree $h$ is smaller than the structural cutoff $h_s=\sqrt{N\mean{h}}$.
While this restricts $p(h,h')$ in \eqref{c1} within the interval $[0,1]$, banning degrees larger than the structural cutoff strongly violates the reality of scale-free networks, where degrees all the way up to the
natural cutoff $(N\mean{h})^{1/(\tau-1)}$ need to be considered. We therefore work with (although many asymptotically equivalent choices are possible; see \cite{hofstad2017b} and Appendix \ref{sec:chcomp})
\begin{equation}\label{c11}
p(h,h')= \min\Big(1,\frac{h h'}{N \mean{h}}\Big),
\end{equation}
putting no further restrictions on the range of the hidden variables (and hence degrees).

In this paper, we shall work with $c(h)$, the local clustering coefficient of a randomly chosen vertex with hidden variable $h$. However, when studying local clustering in real-world data sets, we can only observe $\bar c(k)$, the local clustering coefficient of a vertex of degree $k$. In Appendix \ref{sec:hk} we show that the approximation $\bar{c}(h)\approx c(h)$ is highly accurate. 
We start from the explicit expression for $c(h)$ \cite{boguna2003}, which measures the probability that two randomly chosen edges from $h$ are neighbors, i.e.,
\begin{equation}\label{int1}
c(h)=\int_{h'}\int_{h''} p(h'|h)p(h',h'')p(h''|h)\dd h''\dd h',
\end{equation}
with $p(h'|h)$ the conditional probability that a randomly chosen edge from an $h$-vertex is connected to an $h'$-vertex and $p(h,h')$ as in \eqref{c11}. The goal is now to characterize the $c(h)$-curve (and hence the $\bar c(k)$-curve).

\section{Universal clustering spectrum}\label{sec:three}

\begin{figure}[tb]
	\centering\includegraphics[width=0.4\textwidth]{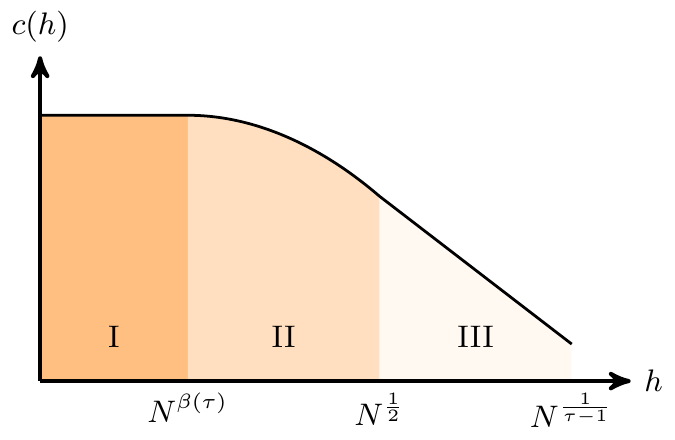}
	\caption{Clustering spectrum $h\mapsto c(h)$ with three different ranges for $h$: the flat range, logarithmic decay, and the power-law decay.}
	\label{fig:curve}
\end{figure}

 The asymptotic evaluation of the double integral \eqref{int1} in the large-$N$ regime reveals three different ranges, defined in terms of the scaling relation between the hidden variable $h$ and the network size $N$. The three ranges together span the entire clustering spectrum as shown in Fig.~\ref{fig:curve}. The detailed calculations are deferred to Appendix \ref{sec:chcomp}.

The first range pertains to the smallest-degree nodes, i.e., vertices with a hidden variable that does not exceed $N^{\beta(\tau)}$ with  $\beta(\tau)= \frac{\tau-2}{\tau-1}$. In this case we show that
\begin{equation}\label{eq:r1}
c(h)\propto N^{2-\tau}\ln N, \quad h\leq N^{\beta(\tau)}.
\end{equation}
In particular, here the local clustering does not depend on the degree and in fact corresponds with the large-$N$ behavior of the global clustering coefficient \cite{hofstad2017b,colomer2012}. Note that the interval $[0,\beta(\tau)]$ diminishes when $\tau$ is close to 2, a possible explanation for why the flat range associated with Range I is hard to recognize in some of the real-world data sets.

Range II considers nodes with hidden variables (degrees) above the threshold $N^{\beta(\tau)}$, but below the structural cutoff $\sqrt{N}$. These nodes start experiencing structural correlations, and close inspection of the integral \eqref{int1} yields
\begin{equation}\label{eq:r2}
c(h)\propto N^{2-\tau}\Big(1+\ln \Big(\frac{\sqrt{N}}{h}\Big)\Big), \quad N^{\beta(\tau)}\leq h \leq \sqrt{N}.
\end{equation}
This range shows relatively slow, logarithmic decay in the clustering spectrum, and is clearly visible in the ten data sets.

Range III considers hidden variables above the structural cutoff, when the restrictive effect of degree-degree correlations becomes more evident. In this range we find that
\begin{equation}\label{eq:r3}
c(h)\propto \frac{1}{N}\Big(\frac{h}{N}\Big)^{-2(3-\tau)}, \quad h\geq \sqrt{N},
\end{equation}
hence power-law decay with a power-law exponent $\alpha=2(3-\tau)$.
Such power-law decay has been observed in many real-world networks
\cite{vazquez2002,ravasz2003,serrano2006b,catanzaro2004, leskovec2008,krioukov2012}, where most networks were found to have the power-law exponent close to one. The asymptotic relation \eqref{eq:r3} shows that the exponent $\alpha$ decreases with $\tau$ and takes values in the entire range $(0,2)$. Table~\ref{tab:data} contains estimated values of $\alpha$ for the ten data sets.

\section{Energy minimization}\label{sec:deep}
\begin{figure}[tb]
	\centering
	\subfloat[]{
		\centering
		\includegraphics[width=0.2\textwidth]{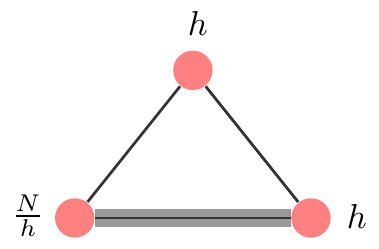}
		\label{fig:contsmall}
	}
\hspace{0.2cm}
	\subfloat[]{
		\centering
		\includegraphics[width=0.2\textwidth]{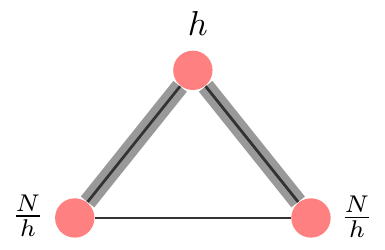}
		\label{fig:contlarge}
	}
	\caption{Orders of magnitude of the major contributions in the different $h$-ranges. The highlighted edges are present with asymptotically positive probability. (a) $h<\sqrt{N}$ (b) $h>\sqrt{N}$.}
	\label{fig:major}
\end{figure}

We now explain why the clustering spectrum splits into three ranges, using an argument that minimizes the energy needed to create triangles among nodes with specific hidden variables.

In all three ranges for $h$, there is one type of `most likely' triangle, as shown in Fig.~\ref{fig:major}. This means that most triangles containing a vertex $v$ with hidden variable $h$ are triangles with two other vertices $v'$ and $v''$ with hidden variables $h'$ and $h''$ of specific sizes, depending on $h$. The probability that a triangle is present between $v$, $v'$ and $v''$ can be written as
\begin{equation}\label{eq:probtr}
\min\left(1,\frac{hh'}{N\mean{h}}\right)\min\left(1,\frac{hh''}{N \mean{h} }\right)\min\left(1,\frac{h'h''}{N\mean{h} }\right).
\end{equation}
While the probability that such a triangle exists among the three nodes thus increases with $h'$ and $h''$, the number of such nodes decreases with
 $h'$ and $h''$ because vertices with higher $h$-values are rarer. Therefore, the maximum contribution to $c(h)$ results from a trade-off between large enough $h',h''$ for likeliness of occurrence of the triangle, and $h',h''$ small enough to have enough copies. Thus, having $h'> N\mean{h}/h$ is not optimal, since then the probability that an edge exists between $v$ and $v'$ no longer increases with $h'$. This results in the bound
\begin{equation}\label{eq:b1}
h',h''\leq \frac{N\mean{h} }{h}.
\end{equation}
Similarly, $h'h''> N\mean{h}$ is also suboptimal, since then further increasing $h'$ and $h''$ does not increase the probability of an edge between $v'$ and $v''$. This gives as a second bound
\begin{equation}\label{eq:b2}
h'h''\leq N\mean{h}.
\end{equation}
In Ranges I and II, $h<\sqrt{N\mean{h}}$, so that $N\mean{h}/h>\sqrt{N\mean{h}}$. In this situation we reach bound~\eqref{eq:b2} before we reach bound~\eqref{eq:b1}. Therefore, the maximum contribution to $c(h)$ comes from $h'h''\approx N$, where also $h',h''<N\mean{h}/h$ because of the bound~\eqref{eq:b1}. Here the probability that the edge between $v'$ and $v''$ exists is large, while the other two edges have a small probability to be present, as shown in Fig.~\ref{fig:contsmall}. Note that for $h$ in Range I, the bound \eqref{eq:b1} is superfluous, since in this regime $N\mean{h}/h>h_c$, while the network does not contain vertices with hidden variables larger than $h_c$. This bound indicates the minimal values of $h'$ such that an $h$-vertex is guaranteed to be connected to an $h'$-vertex. Thus, vertices in Range I are not even guaranteed to have connections to the highest degree vertices, hence they are not affected by the single-edge constraints. Therefore the value of $c(h)$ in Range I is independent of $h$.

In Range III, $h>\sqrt{N\mean{h}}$, so that $N\mean{h}/h<\sqrt{N\mean{h}}$. Therefore, we reach bound~\eqref{eq:b1} before we reach bound~\eqref{eq:b2}. Thus, we maximize the contribution to the number of triangles by choosing $h',h''\approx N\mean{h}/h$. Then the probability that the edge from $v$ to $v'$ and from $v$ to $v''$ is present is large, while the probability that the edge between $v'$ and $v''$ exists is small, as illustrated in Fig.~\ref{fig:contlarge}.

\section{Convergence rate}\label{sec:deep2}

We next ask  how large networks should be, or become, before they reveal the features of the universal clustering spectrum.
In other words, while the results in this paper are shown for the large-$N$ limit, for what finite $N$-values can we expect to see the different ranges and clustering decay?
To bring networks of different sizes $N$ on a comparable footing, we consider
\begin{equation}\label{eq:sigmam}
\sigma_N(t)=\frac{\ln\left(c(h)/c(h_c)\right)}{\ln(N\mean{h})}, \quad h=(N\mean{h})^t,
\end{equation}
for $0\leq t \leq \tfrac{1}{\tau-1}$. The slope of $\sigma_N(t)$ can be interpreted as a measure of the decay of $c(h)$ at $h=(N\mean{h})^t$, and all curves share the same right end of the spectrum; see Appendix~\ref{sec:hcder} for more details.
Figure~\ref{fig:chfinite} shows this rescaled clustering spectrum for synthetic networks generated with the hidden variable model with $\tau=2.25$. Already $10^4$ vertices reveal the essential features of the spectrum: the decay and the three ranges. Increasing the network size further to $10^5$ and $10^6$ nodes shows that the spectrum settles on the limiting curve. Here we note that the real-world networks reported in Figs.~\ref{fig:chas} and~\ref{fig:ch} are also of order $10^5$-$10^6$ nodes, see Table~\ref{tab:data}.

\begin{figure}[ht]
	\centering\includegraphics[width=0.4\textwidth]{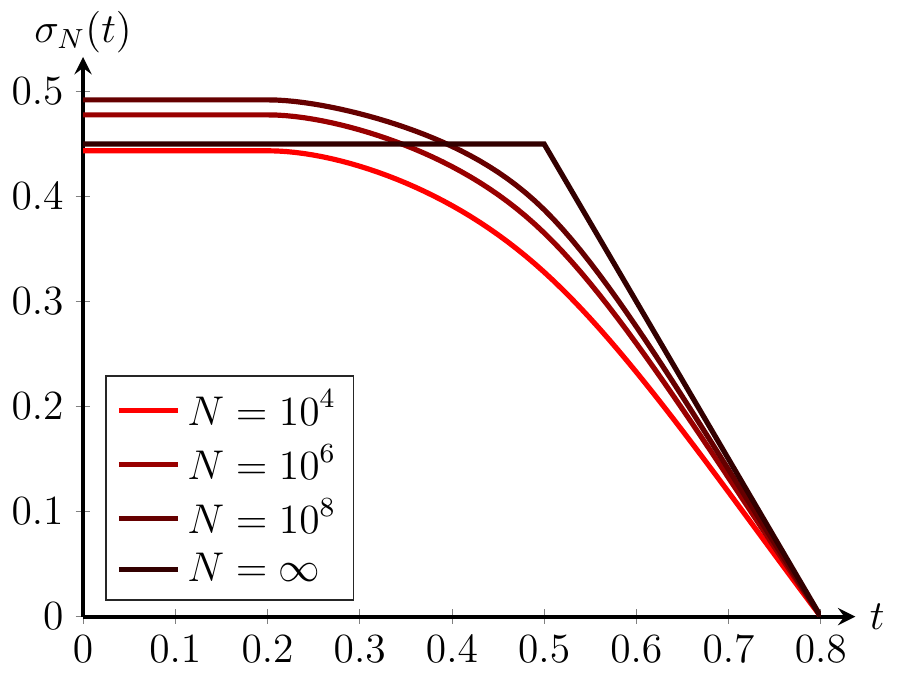}
	\caption{$\sigma_N(t)$ for $N=10^4,10^6$ and $10^8$ together with the limiting function, using $\tau=2.25$, for which $\tfrac{1}{\tau-1}=0.8$. } 
	\label{fig:chfinite}
\end{figure}
\begin{table}[htbp]
	\centering
	\begin{ruledtabular}
		\begin{tabular}{lrrrr}
		               & $N$  &                $\tau$ & g.o.f. &$\alpha$ \\ 
		\textbf{Hudong }         &               1.984.484 &                 2,30 &   0.00  &0,85 \\
		\textbf{Baidu}           &               2.141.300 &                   2,29 &   0.00 & 0,80 \\
		\textbf{Wordnet}         &                146.005 &             2,47 & 0.00    &1,01 \\
		\textbf{Google web}      &                875.713 &                2,73 & 0.00  &  1,03 \\
		\textbf{AS-Skitter}      &               1.696.415 &           2,35 &  0.06  & 1,12 \\
		\textbf{TREC-WT10g}      &               1.601.787 &          2,23 & 0.00  &  0,99 \\
		\textbf{Wiki-talk}       &               2.394.385 &            2,46 &  0.00   &1,54 \\
		\textbf{Catster/Dogster} &                623.766 &            2,13 &    0.00 & 1,20 \\
		\textbf{Gowalla}         &                196.591 &          2,65 &    0.80 &1,24 \\
		\textbf{Youtube}         &               1.134.890 & 		2,22&  0.00   &1,05
		\end{tabular}
	\end{ruledtabular}%
	\caption{Data sets. $N$ denotes the number of vertices, $\tau$ the exponent of the tail of the degree distribution estimated by the method proposed in~\cite{clauset2009} together with the goodness of fit criterion proposed in~\cite{clauset2009} (when the goodness of fit is at least 0.10, a power-law tail cannot be rejected), and $\alpha$ denotes the exponent of $c(k)$.}
	\label{tab:data}%
\end{table}%
Figure~\ref{fig:chfinite} also brings to bear a potential pitfall when the goal is to obtain statistically accurate estimates for the slope of $c(h)$. Observe the extremely slow convergence to the limiting curve for $N=\infty$; a well documented property of certain clustering measures~\cite{boguna2009,colomer2012,janssen2015,hofstad2017b}. In Appendix~\ref{sec:hcder}
we again use the integral expression \eqref{int1} to characterize the limiting curve for $N=\infty$ and
the rate of convergence as function of $N$, and indeed extreme $N$-values are required for statistically reliable slope estimates for e.g.~$t$-values of $\tfrac12$ and $\tfrac{1}{\tau-1}$; this is also apparent from visual inspection of Fig.~\ref{fig:chfinite}.
Therefore, the estimates in Table~\ref{tab:data} only serve as indicative values of $\alpha$. Finally, observe that Range II disappears in the limiting curve, due to the rescaling in \eqref{eq:sigmam}, but again only for extreme $N$-values. Because this paper is about structure rather than statistical estimation, the slow convergence in fact provides additional support for the persistence of Range II in Figs.~\ref{fig:chas} and~\ref{fig:ch}.

Table~\ref{tab:data} also shows that the relation $\alpha=-2(3-\tau)$ is inaccurate for the real-world data sets, in turn affecting the theoretical boundaries of the three regimes  indicated in Fig.~\ref{fig:ch}. One explanation for this inaccuracy is that the real-world networks might not follow pure power-law distributions, as measured by the goodness of fit criterion in Table~\ref{tab:data}, and visualized in Appendix~\ref{sec:degree}. Furthermore, real-world networks are usually highly clustered and contain community structures, whereas the hidden variable model is locally tree-like. These modular structure may explain, for example, why the power-law decay of the hidden variable model is less pronounced in the three social networks of Fig.~\ref{fig:ch}. It is remarkable that despite these differences between hidden variable models and real-world networks, the global shape of the $c(k)$ curve of the hidden variable model is still visible in these heavy-tailed real-world networks.

\section{Discussion}\label{sec:disc}

The hidden variable model gives rise to {\it single-edge} networks in which pairs of vertices can only be connected once. Hierarchical modularity and the decaying clustering spectrum have been contributed to this restriction that no two vertices have more than one edge connecting them~\cite{pastor2001b, maslov2004,park2003,newman2002assortative,newman2003}. The physical intuition is that the single-edge constraint leads to far fewer connections between high-degree vertices than anticipated based on randomly assigned edges. 
We have indeed confirmed this intuition, not only through analytically revealing the universal clustering curve, but also by providing an alternative derivation of the three ranges based on energy minimization and structural correlations.

We now show that the clustering spectrum revealed using the hidden variable model, also appears for a second widely studied null model. This second model cannot be the Configuration Model (CM), which preserves the degree distribution by making connections between vertices in the most random way possible~\cite{bollobas1980, newman2001}. Indeed, because of the random edge assignment, the CM has no degree correlations, leading in the case of scale-free networks with diverging second moment to uncorrelated networks with non-negligible fractions of self-loops (a vertex joined to itself) and multiple connections (two vertices connected by more than one edge). This picture changes dramatically when self-loops and multiple edges are avoided,
a restriction mostly felt by the high-degree nodes, who can no longer establish multiple edges among each other. 

\begin{figure}[t]
	\centering
	\includegraphics[width=0.4\textwidth]{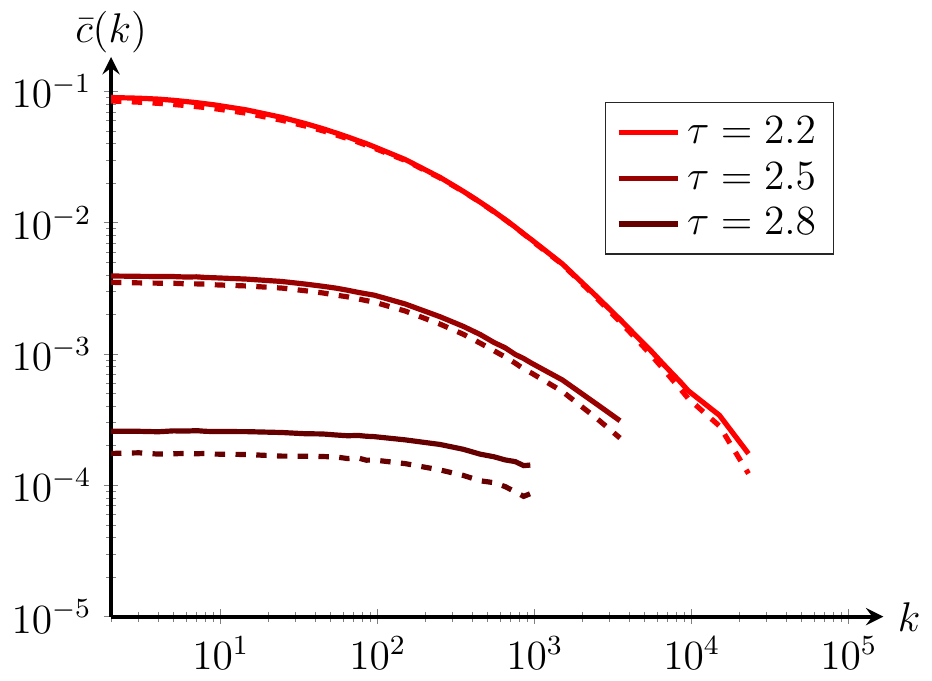}
	\caption{$\bar c(k)$ for a hidden variable model with connection probabilities~\eqref{eq:conecm} (solid line) and an erased configuration model (dashed line). The presented values of $\bar c(k)$ are averages over $10^4$ realizations of networks of size $N=10^5$.  }
	\label{fig:ECMhidden}
\end{figure}

We therefore consider the Erased Configuration Model (ECM) that takes a sample from the CM and then erases all the self-loops and multiple edges. While this removes some of the edges in the graph, thus violating the hard constraint, only a small proportion of the edges is removed, so that the degree of vertex $j$ in ECM is still close to $D_j$~\cite[Chapter 7]{hofstad2009}. In the ECM, the probability that a vertex with degree $D_i$ is connected to a vertex with degree $D_j$ can be approximated by $1-\me^{-D_iD_j/\mean{D}N}$~\cite[Eq.(4.9)]{hofstad2005}. Therefore, we expect the ECM and the hidden variable model to have similar properties (see e.g.~\cite{hofstad2017b}) when we choose
\begin{equation}\label{eq:conecm}
p(h,h')= 1-\me^{-h h'/N\mean{h}}\approx\frac{h h'}{N\mean{h}}.
\end{equation}
Figure~\ref{fig:ECMhidden} illustrates how both null models generate highly similar spectra, which provides additional support for the claim that the clustering spectrum is a universal property of simple scale-free networks. The ECM is more difficult to deal with compared to hidden variable models, since edges in ECM are not independent. In particular, we expect that these dependencies vanish for the $k\mapsto\bar{c}(k)$ curve. Establishing the universality of the $k\mapsto \bar c(k)$ curve for other random graph null models such as ECM,  networks with an underlying geometric space~\cite{serrano2008} or hierarchical configuration models~\cite{stegehuis2015} is a major research direction.
The ECM and the hidden variable model are both null models with soft constraints on the degrees. Putting hard constraints on the degrees with the CM, has the nice property that simple graphs generated using this null model are uniform samples of all simple graphs with the same degree sequence. Dealing with such uniform samples is notoriously hard when the second moment of the degrees is diverging, for example since the CM will yield many edges between high-degree vertices. This makes sampling uniform graphs difficult~\cite{milo2003, viger2005,delgenio2010}. Thus, the joint requirement of hard degree and single-edge constraints, as in the CM, presents formidable technical challenges. Whether our results for the $k\mapsto\bar{c}(k)$ curve for soft-constraint models also carry over to these uniform simple graphs is a challenging open problem.

In this paper we have investigated the presence of triangles in the hidden variable model.
 We have shown that by first conditioning on the node degree, there arises a unique
 `most likely' triangle with two other vertices of specific degrees. We have not only explained this insight heuristically, but it is also reflected in the elaborate analysis of the double integral for $c(h)$ in Appendix \ref{sec:chcomp}. As such, we have introduced an intuitive and tractable mathematical method for asymptotic triangle counting. It is likely that the method carries over to counting other motifs, such as squares, or complete graphs of larger sizes. For any given motif, and first conditioning on the node degree, we again expect to find specific configuration that are most likely. Further mathematical challenges need to be overcome, though, because we expect that the `most likely' configurations critically depend on the precise motif topologies and the associated energy minimization problems.

\acknowledgements
This work is supported by NWO TOP grant 613.001.451 and by the NWO Gravitation Networks grant 024.002.003.
The work of RvdH is further supported by the NWO VICI grant 639.033.806.  The work of JvL is further supported by an NWO TOP-GO grant and by an ERC Starting Grant.

\appendix

\section{Derivation for the three ranges}\label{sec:chcomp}
In this appendix, we compute $c(h)$ in~\eqref{int1}, and we show that $c(h)$ can be approximated by~\eqref{eq:r1},~\eqref{eq:r2}, or~\eqref{eq:r3}, depending on the value of $h$.
 Throughout the appendix, we assume that $p(h, h')=\min(1,h h'/\hst)$ and  $\rho(h)=Ch^{-\tau}$. Then, the derivation of $c(h)$ in~\cite{colomer2013} yields
\begin{align}\label{eq:ch}
&c(\w)= \frac{\int_{1}^{h_c}\int_{1}^{h_c}\rho(\w')p(\w,\w')\rho(\w'')p(\w,\w'')p(\w',\w'')\dd \w''\dd \w'}{\big[\int_{1}^{h_c}\rho(\w')p(\w,\w')\dd \w'\big]^2}\nonumber\\
& = \frac{\int_{1}^{h_c}\int_{1}^{h_c}(\w'\w '')^{-\tau}\min(\frac{\w\w'}{\hst}, 1)\min(\frac{\w\w''}{\hst}, 1)\min(\frac{\w'\w''}{\hst}, 1)\dd \w''\dd \w'}{\big[\int_{1}^{h_c}(\w')^{-\tau}\min(\frac{\w\w'}{\hst}, 1)\dd \w'\big]^2}.
\end{align}
Computing $c(h)$ will also allow us to compute
\begin{equation}\label{eq:sigma}
\sigma_N(t)=\frac{\ln(c(h)/c(h_{\text{ref}}))}{\ln(N\mean{h})},\quad h=(N\mean{h})^t,
\end{equation}
for $0\leq t\leq \tfrac{1}{\tau-1}$, where $h_{\text{ref}}\in[0,h_c]$ is fixed.
We are interested in computing the value of $\sigma_N(t)$ for large values of $N$.

Adopting the standard choices \cite{hofstad2017b}
\begin{equation}
\hs=\sqrt{N\mean{h}},\quad h_c=(N\mean{h})^{1/(\tau-1)},
\end{equation}
and setting $h_{\min}=1$ gives
\begin{equation}
\mean{h}=\frac{\tau-1}{\tau-2}\frac{1-N^{2-\tau}}{1-N^{1-\tau}}.
\end{equation}
For ease of notation in the proofs below, we will use
\begin{equation}\label{eq:ab}
a=\hs^{-1}=(N\mean{h})^{-1/2},\quad b=\frac{h_c}{\hs}=(N\mean{h})^{\frac{3-\tau}{2(\tau-1)}},
\end{equation}
and
\begin{equation}\label{eq:r}
r(u)=\min(u,1).
\end{equation}
In this notation,~\eqref{eq:ch} can be succinctly written as
\begin{equation}\label{eq:chab}
c(h)=\frac{\int_{a }^{b}\int_{a }^{b} (xy)^{-\tau}r(a \w x)r(a\w y)r(xy)\dd x\dd y}{\big[\int_{a  }^{b}x^{-\tau}r(a \w x)\dd x\big]^2}.
\end{equation}
Because of the four min operators in the expression~\eqref{eq:ch}, we have to consider various $h$-ranges. We compute the value of $c(h)$ in these three ranges one by one.

\paragraph*{Range I: $h<\hst/h_c$.}
We now show that in this range
\begin{equation}\label{eq:chapprsmall}
c(h)\approx \frac{\tau-2}{3-\tau}\hs^{4-2\tau}\ln\Big(\frac{h_c^2}{\hst}\Big)\propto N^{2-\tau}\ln N,
\end{equation}
which proves~\eqref{eq:r1}.

This range corresponds to $h<1/(ab)$ with $a$ and $b$ as in~\eqref{eq:ab}. In this range, $r(ahx)=ahx$ and $r(ahy)=ahy$ for all $x\in[a,b]$.
This yields for $c(\w)$
\begin{equation}\label{eq:ckeq}
\begin{aligned}[b]
c(\w)& = \frac{\int_{a}^{b}\int_{a}^{b}(xy)^{1-\tau}r(xy)\dd x\dd y}{\big[\int_{a}^{b}x^{1-\tau}\dd x \big]^2}.
\end{aligned}
\end{equation}
For the denominator we compute
\begin{equation}\label{eq:denomchsmall}
\int_{a}^{b}x^{1-\tau}\dd x = \frac{a^{2-\tau}-b^{2-\tau}}{\tau-2}.
\end{equation}
Since $a\ll b$, this can be approximated as
\begin{equation}
\frac{a^{2-\tau}-b^{2-\tau}}{\tau-2}\approx \frac{a^{2-\tau}}{\tau-2}.
\end{equation}
We can compute the numerator of~\eqref{eq:ckeq} as
\begin{equation}\label{eq:chsmall}
\begin{aligned}[b]
& \int_{a}^{b}\int_{a}^{b}(xy)^{1-\tau}r(xy)\dd x\dd y  \\
& = \int_{a}^{1/b}\int_{a}^{b}{(xy)^{2-\tau}}\dd x\dd y+ \int_{1/b}^{b}\int_{a}^{1/x}{(xy)^{2-\tau}}\dd x\dd y \\
&\quad + \int_{1/b}^{b}\int_{1/x}^{b}(xy)^{1-\tau}\dd x\dd y\\
& = \frac{\left(b^{\tau-3}-a^{3-\tau}\right)\left(b^{3-\tau}-a^{3-\tau}\right)}{(3-\tau)^2}\\
&\quad  + \frac{1}{3-\tau}\left(\ln\left(b^2\right)-\frac{a^{3-\tau}\left(b^{3-\tau}-b^{\tau-3}\right)}{3-\tau}\right)\\
& \quad +\frac{1}{2-\tau}\left(\frac{b^{2-\tau}(b^{2-\tau}-b^{\tau-2})}{2-\tau}-\ln\left(b^2\right)\right)\\
& =\frac{\ln\left(b^2\right)}{(3-\tau)(\tau-2)}- \frac{1-b^{4-2\tau}}{(\tau-2)^2} +\frac{1-2(ab)^{3-\tau} +a^{6-2\tau}}{(3-\tau)^2}.
\end{aligned}
\end{equation}
The first of these three terms dominates when
\begin{equation}\label{eq:Lndom1}
\frac{3-\tau}{\tau-1}\frac{\ln(N\mean{h})}{(3-\tau)(\tau-2)}\gg \frac{1}{(\tau-2)^2}
\end{equation}
and
\begin{equation}\label{eq:Lndom2}
\frac{3-\tau}{\tau-1}\frac{\ln(N\mean{h})}{(3-\tau)(\tau-2)}\gg \frac{1}{(3-\tau)^2},
\end{equation}
where we have used that $b^2=(N\mean{h})^{(3-\tau)/(\tau-1)}$.
Thus, when $\ln(N\mean{h} )$ is large compared to $(\tau-1)/(\tau-2)$ and $(\tau-1)(\tau-2)/(\tau-3)^2$, we obtain
\begin{equation}
c(h)\approx \frac{\tau-2}{3-\tau}a^{2\tau-4}\ln\left(b^2\right)\propto N^{2-\tau}\ln(N),
\end{equation}
which proves~\eqref{eq:chapprsmall}.

\paragraph*{Range II: $\hst/h_c<h<\hs$}
In this range, we show that
\begin{equation}\label{eq:chapprmiddle}
c(h)\approx \hs^{4-2\tau}\frac{\ln\left(\frac{\hst}{h^2}\right)+M}{(\tau-2)(3-\tau)}\propto N^{2-\tau}\left(\ln\left(N/h^2\right)+M\right),
\end{equation}
for some positive constant $M$, which proves~\eqref{eq:r2}.

This range corresponds to $(ab)^{-1}<h<a^{-1}$. For these values of $h$, we have $ahx,ahy=1$ for $x,y=(ah)^{-1}\in(1,b)$ and $xy=1$ for $y=1/x\in[a,b]$ when $b^{-1}<x<b$.
Then for the denominator of~\eqref{eq:chab} we compute
\begin{equation}\label{eq:numchsmall}
\begin{aligned}[b]
&  \int_{a}^{1/(ah)}ahx^{1-\tau}\dd x+ \int_{1/(ah)}^{b}x^{-\tau}\dd x \\
 & = \frac{1}{\tau-2}(a^{3-\tau}h-(ah)^{\tau-1})\\
 &\quad +\frac{1}{\tau-1}((ah)^{\tau-1}-b^{1-\tau})	\\
& = ah\left(\frac{a^{2-\tau}}{\tau-2}-\frac{(ah)^{\tau-2}}{(\tau-1)(\tau-2)}-\frac{b^{1-\tau}/(ah)}{\tau-1}\right).
\end{aligned}
\end{equation}

Splitting up the integral in the numerator results in
\begin{equation}
\begin{aligned}[b]
& \text{Num}(h)=\int_{a }^{b}\int_{a }^{b} (xy)^{-\tau}r(a \w x)r(a\w y)r(xy)\dd x\dd y\\
& =\int_{1/(ah)}^{b}\int_{1/(ah)}^{b}(xy)^{-\tau}\dd y\dd x \\
& \quad +{2a h}\int_{1/(ah)}^{b}\int_{1/x}^{1/(ah)}(xy)^{-\tau}y\dd y\dd x\\
& \quad+ {2 ah}\int_{1/(ah)}^{b}\int_a^{1/x}(xy)^{1-\tau}y\dd y\dd x\\
&\quad + { a^2h^2}\int_{ah}^{1/(ah)}\int_a^{1/x}(xy)^{2-\tau}\dd y\dd x\\
& \quad+ { a^2h^2}\int_{ah}^{1/(ah)}\int_{1/x}^{1/(ah)}(xy)^{1-\tau}\dd y\dd x \\
&\quad + {a^2 h^2}\int_a^{ah}\int_{a}^{1/(ah)}(xy)^{2-\tau}\dd y\dd x\\
&=:I_1+I_2+I_3+I_4+I_5+I_6,
\end{aligned}
\end{equation}
where the factors 2 arise by symmetry of the integrand in $x$ and $y$. Computing these integrals yields
\begin{align}
I_1&=a^2h^2\left(\frac{(ah)^{\tau-2}-a^{-1}b^{1-\tau}h^{-1}}{\tau-1}\right)^2,\label{eq:I1middle}\\
I_2&=2a^2h^2\Big(\frac{1-1/(abh)}{\tau-2}-\frac{(ah)^{2\tau-4}}{(\tau-1)(\tau-2)}\\
& \quad \times\left(1-(abh)^{1-\tau}\right)\Big),\\
I_3&=2a^2h^2\Big(\frac{1-1/(abh)}{3-\tau}-\frac{h^{\tau-3}\left(1-(abh)^{2-\tau}\right)}{(3-\tau)(\tau-2)}\Big),\\
I_4&=a^2h^2\left(\frac{\ln((ah)^{-2})}{3-\tau}+\frac{(a^2h)^{3-\tau}-h^{\tau-3}}{(3-\tau)^2}\right),\\
I_5&=a^2h^2\left(\frac{\ln((ah)^{-2})}{\tau-2}-\frac{1-(ah)^{2\tau-4}}{(\tau-2)^2}\right),\\
I_6&=a^2h^{2}\left(\frac{1-h^{\tau-3}+a^{6-2\tau}-(a^2h)^{3-\tau}}{(3-\tau)^2}\right)\label{eq:Iendmiddle}.
\end{align}
We have $ah<1<ahb$ and so the leading behavior of $\text{Num}(h)$ is determined by the terms involving $\ln((ah)^{-2})$ in $I_3$ and $I_4$, all other terms being bounded. Retaining only these dominant terms, we get
\begin{equation}
\text{Num}(h)=a^2h^2\frac{\ln((ah)^{-2})}{(\tau-2)(3-\tau)}(1+o(1)),
\end{equation}
provided that $ah\to 0$ as $N\to \infty$. In terms of the variable $t$ in $h=(N\mean{h})^t$, see~\eqref{eq:sigmam} and~\eqref{eq:sigma}, this condition holds when we restrict to $t\in[(\tau-2)/(\tau-1),\tfrac12-\varepsilon]$ for any $\varepsilon>0$. Furthermore, from~\eqref{eq:numchsmall},
\begin{equation}\label{eq:numapprmiddle}
\left(\int_{a}^{b}x^{-\tau}r(ahx)\dd x\right)^2 = a^2h^2\left(\frac{a^{2-\tau}}{\tau-2}\right)^2(1+o(1)).
\end{equation}
Hence, when $ah\to 0$, we have
\begin{equation}\label{eq:chmiddlegood}
c(h)= \frac{\tau-2}{3-\tau}a^{2\tau-4}\ln\left((ah)^{-2}\right)(1+o(1))\propto N^{2-\tau}\ln\left(N/h^2\right).
\end{equation}

We compute $c(h=1/a)$ asymptotically by retaining only all constant terms between brackets in~\eqref{eq:I1middle}-\eqref{eq:Iendmiddle} since all other terms vanish or tend to 0 as $N\to\infty$. This gives
\begin{equation}
\begin{aligned}[b]
&\text{Num}(h=1/a)=a^2h^2\Big(\frac{1}{(\tau-1)^2}+\frac{2}{\tau-2}\\
&\quad -\frac{2}{(\tau-1)(\tau-2)}+\frac{2}{3-\tau}+\frac{1}{(3-\tau)^2}\Big)(1+o(1))\\
& = Pa^2h^2(1+o(1)),
\end{aligned}
\end{equation}
where $P=\frac{1}{(\tau-1)^2}+\frac{1}{(3-\tau)^2}+\frac{2}{\tau-1}+\frac{2}{3-\tau}$. Together with~\eqref{eq:numapprmiddle}, we find
\begin{equation}\label{eq:chbound}
c(h=1/a)=P(\tau-2)^2a^{2\tau-4}(1+o(1))\propto N^{2-\tau.}
\end{equation}
In~\cite{hofstad2017b}, it has been shown that $c(h)$ decreases in $h$, and then~\eqref{eq:chapprmiddle} follows from~\eqref{eq:chmiddlegood} and~\eqref{eq:chbound}.

\paragraph*{Range III: $\hs<h<h_c$.}
We now show that when $\hs<h<h_c$, then
\begin{equation}\label{eq:chapprlarge}
c(h)\approx \frac{1}{(3-\tau)^2}(\hs/h)^{6-2\tau}\hs^{4-2\tau}\propto N^{5-2\tau}h^{2\tau-6},
\end{equation}
which proves~\eqref{eq:r3}.

This range corresponds to $1/a<h<b/a$. The denominator of~\eqref{eq:chab} remains the same as in the previous range and is given by~\eqref{eq:numchsmall}. Splitting up the integral in the numerator of~\eqref{eq:chab} now results in
\begin{equation}
\begin{aligned}[b]
&\text{Num}(h)=
\int_{a }^{b}\int_{a }^{b} (xy)^{-\tau}r(a \w x)r(a\w y)r(xy)\dd x\dd y\\
 &= \int_{1/(ah)}^{ah}\int_{1/x}^{b}(xy)^{-\tau}\dd y\dd x+\int_{ah}^{b}\int_{1/(ah)}^{b}(xy)^{-\tau}\dd y\dd x\\
& \quad +\int_{1/(ah)}^{ah}\int_{1/(ah)}^{1/x}(xy)^{1-\tau}\dd y\dd x \\
&\quad +{2ah}\int_{ah}^{b}\int_{1/x}^{1/(ah)}(xy)^{-\tau}y\dd y\dd x\\
& \quad +{ 2ah}\int_{1/(ah)}^{ah}\int_a^{1/(ah)}(xy)^{1-\tau}y\dd y\dd x \\
&\quad +{2ah}\int_{ah}^{b}\int_a^{1/x}(xy)^{1-\tau}y\dd y\dd x\\
& \quad + a^2 h^2\int_a^{1/(ah)}\int_a^{1/(ah)}(xy)^{2-\tau}\dd y\dd x\\
&=:I_1+I_2+I_3+I_4+I_5+I_6+I_7.
\end{aligned}
\end{equation}
Computing these integrals yields
\begin{align}
I_1&=a^2h^2\Big((ah)^{-2}\frac{\ln(a^2h^2)}{\tau-1}\\
&\quad +\frac{b^{1-\tau}\left((ah)^{-\tau-1}-(ah)^{\tau-3}\right)}{(\tau-1)^2}\Big),\label{eq:I1large}\\
I_2& = a^2h^2\Bigg(\frac{(ah)^{-2}+b^{2-2\tau}(ah)^{-2}}{(\tau-1)^2}\\
&\quad -\frac{b^{1-\tau}\left((ah)^{\tau-3} +(ah)^{-\tau-1}\right)}{(\tau-1)^2}\Bigg),\\
I_3&=a^2h^2\left(-(ah)^{-2}\frac{\ln(a^2h^2)}{\tau-2}+\frac{(ah)^{2\tau-6}-(ah)^{-2}}{(\tau-2)^2}\right),\\
I_4 &=2a^2h^{-2}\left(-\frac{(abh)^{-1}}{\tau-2}+\frac{(ah)^{-2}}{\tau-1}+\frac{b^{1-\tau}(ah)^{\tau-3}}{(\tau-1)(\tau-2)}\right),\\
I_5&=2a^2h^2\left(\frac{(ah)^{2\tau-6}+h^{1-\tau}a^{4-2\tau}-h^{\tau-3}-(ah)^{-2}}{(3-\tau)(\tau-2)}\right),\\
I_6&=2a^2h^2\Big(\frac{(ab)^{2-\tau}h^{-1}-h^{1-\tau}a^{4-2\tau}}{(3-\tau)(\tau-2)}\\
& \quad  -\frac{(abh)^{-1}-(ah)^{-2}}{3-\tau}\Big),\\
I_7&=a^2h^2\left(\frac{a^{6-2\tau}-2h^{\tau-3}+(ah)^{2\tau-6}}{\tau-3}\right).\label{eq:Iendlarge}
\end{align}
A careful inspection of the terms between brackets in~\eqref{eq:I1large} and~\eqref{eq:Iendlarge} shows that the terms involving $(ah)^{2\tau-6}$ are dominant when $ah\to\infty$. In terms of the variable $t$ in $h=(N\mean{h})^t$, see~\eqref{eq:sigmam} and~\eqref{eq:sigma}, we have that $ah\to\infty$ when we restrict to $t\in[\tfrac{1}{2}+\varepsilon,1/(\tau-1)]$ for any $\varepsilon>0$. When we retain only these dominant terms, we have, when $ah\to\infty$,
\begin{equation}
\begin{aligned}[b]
&\text{Num}(h)=a^2h^2(ah)^{2\tau-6}\\
&\times \left(\frac{1}{(\tau-2)^2}+\frac{2}{(3-\tau)(\tau-2)}+\frac{1}{(3-\tau)^2}\right)(1+o(1))\\
& = a^2h^2\frac{(ah)^{2\tau-6}}{(\tau-2)^2(3-\tau)^2}(1+o(1)).
\end{aligned}
\end{equation}
Using~\eqref{eq:numapprmiddle} again, we get, when $ah\to\infty$,
\begin{equation}
c(h)= \frac{1}{(3-\tau)^2}(ah)^{2\tau-6}a^{2\tau-4}(1+o(1))\propto N^{5-2\tau}h^{2\tau-6}.
\end{equation}
Furthermore, $c(1/a)$ is given by~\eqref{eq:chbound}, while $c(h)$ decreases in $h$. This gives~\eqref{eq:chapprlarge}.

\paragraph*{Other connection probabilities}
\begin{figure}[tb]
	\centering
	\includegraphics[width=0.4\textwidth]{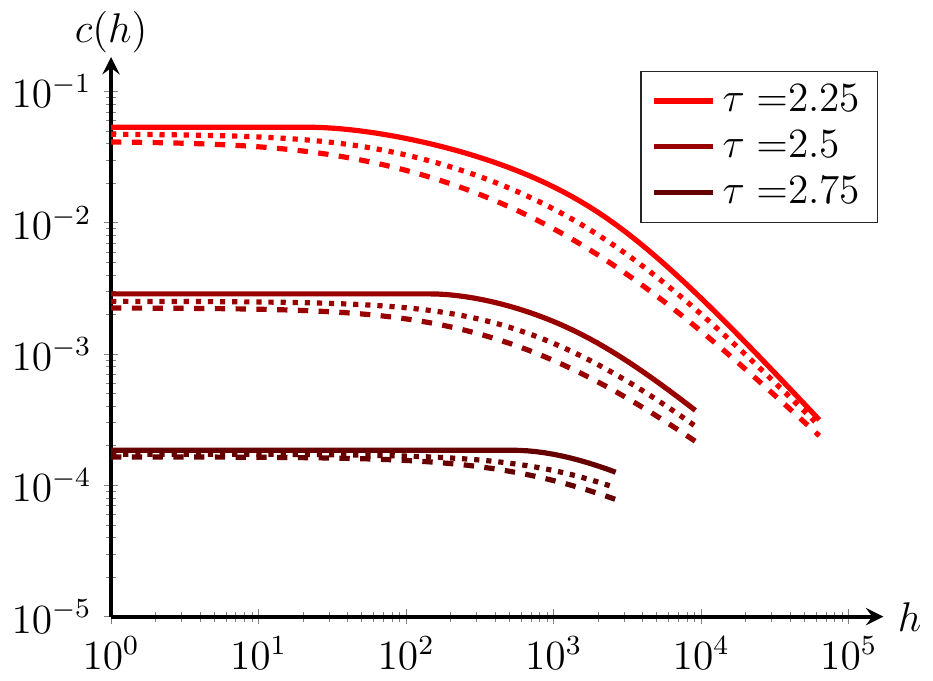}
	\caption{$c(h)$ for $r(u)=\min(u,1)$ (line), $r(u)=u/(1+u)$ (dashed) and $r(u)=1-{\rm e}^{-u}$ (dotted), obtained by calculating~\eqref{eq:chab} numerically.}
	\label{fig:ru}
\end{figure}
In~\cite{hofstad2017b} we have presented a class of functions $r(u)=uf(u)$, $u\geq 0$, so that
\begin{equation}\label{c2x}
p(h,h')=r(u)\quad {\rm with} \quad u=\frac{h h'}{h_s^2}.
\end{equation}
has appropriate monotonicity properties. The maximal member $r(u)=\min(u,1)$ of this class yields $p$ in~\eqref{c11} and is quite representative of the whole class, while allowing explicit computation and asymptotic analysis of $c(h)$ as in \cite{hofstad2017b} and this paper.
Figure \ref{fig:ru} shows that other asymptotically equivalent choices such as
 $r(u)=u/(1+u)$ and $r(u)=1-{\rm e}^{-u}$ have comparable clustering spectra. A minor difference is that the choice $r(u)=\min(1,u)$ for $p$ in~\eqref{c11} forces $c(h)$ to be constant on the range $h\leq N^{\beta(\tau)}$, while the other two choices show a gentle decrease.

\paragraph*{Limiting form of $\sigma_N(t)$ and finite-size effects}
We consider $\sigma_N(t)$ as in~\eqref{eq:sigma} with $h_{\text{ref}}=0$. Using~\eqref{eq:chapprsmall},~\eqref{eq:chapprmiddle} and~\eqref{eq:chapprlarge}, it is readily seen that
\begin{equation}\label{eq:sigmalim}
\lim_{N\to\infty}\sigma_N(t)=\begin{cases}
0,&\quad 0\leq t\leq \tfrac{1}{2},\\
(3-\tau)(1-2t) ,&\quad \tfrac{1}{2}\leq t \leq \tfrac{1}{\tau-1}.
\end{cases}
\end{equation}
Hence, some of the detailed information that is present in~\eqref{eq:chapprsmall},~\eqref{eq:chapprmiddle} and~\eqref{eq:chapprlarge}, disappears when taking the limit as in~\eqref{eq:sigmalim}. This is in particular so for the $\ln N$-factor in~\eqref{eq:chapprsmall} and the logarithmic decaying factor $\ln(N^2/h)$ in Region II.

Consider $\sigma_N(t)$ of~\eqref{eq:sigma} with $h_{\text{ref}}=h_c$ as is done in Fig.~\ref{fig:chfinite}. It follows from the detailed form of~\eqref{eq:chapprsmall} and~\eqref{eq:chapprlarge}, that
\begin{equation}\label{eq:sigmaappr}
\sigma_N(0)=\frac{\ln(c(0)/c(h_c))}{\ln(N\mean{h})}=\gamma+\frac{\ln(\beta y)}{y},
\end{equation}
where
\begin{equation}
\gamma=\frac{(3-\tau)^2}{\tau-1},\quad \beta =(\tau-2)\gamma, \quad y=\ln(N\mean{h}).
\end{equation}
We have that $\sigma_N(0)\to\gamma$ as $N\to\infty$, and the right-hand side of~\eqref{eq:sigmaappr} exceeds this limit $\gamma$ from $y=1/\beta$ onwards with a maximum excess $\beta/\me$ for $N\mean{h}$ as large as $\exp(\me/\beta)$. This explains why the excess of $\sigma_N(0)$ over its limit value in Fig.~\ref{fig:chfinite} with $\me^{\me/\beta}=3\times 10^{10}$ when $\tau=9/4$ persists.

\section{Exact and asymptotic result for decay rate of $c(h)$ at $h=h_c$ and $h=\hs$}\label{sec:hcder}
We let $h_c=(N\mean{h})^{1/(\tau-1)}$, where we assume that $N$ is so large that $h_c\leq N$. This requires $N$ to be of the order $(1/\varepsilon)^{1/\varepsilon}$, where $\varepsilon=\tau-2$. We again consider the function $\sigma_N(t)$ of~\eqref{eq:sigmam},
\begin{equation}
\sigma_N(t)=\frac{\ln(c(h)/c(h_{\text{ref}}))}{\ln(N\mean{h})},\quad h=(N\mean{h})^t,
\end{equation}
for $0\leq t\leq \tfrac{1}{\tau-1}$ and $h_{\text{ref}}$ is fixed, so that
\begin{equation}
c(h)=c(h_{\text{ref}})(N\mean{h})^{\sigma_N(t)},\quad  h=(N\mean{h})^t.
\end{equation}
When we fix a $t_0$ and linearize $\sigma_N(t)$ around $t_0$, we get
\begin{equation}
\begin{aligned}[b]
c(h)& \approx c(h_{\text{ref}})(N\mean{h})^{\sigma_N(t_0)+(t-t_0)\sigma_N'(t_0)} \\
& =c(h_{0})\left(\frac{h}{h_0}\right)^{\sigma_N'(t_0)}
\end{aligned}
\end{equation}
so that $\sigma_N'(t)=\frac{\text{d}}{\text{dt}}\sigma_N(t)$ is a measure for the decay rate of $c(h)$ at $h= h_0=(N\mean{h})^{t_0}$.

In this appendix, we compute an exact expression for $\sigma_N'(t)$ at $t=\tfrac{1}{\tau-1}$, we compute its limit as $N\to\infty$ and discuss convergence speed, and we show that this limit is a lower bound for $\sigma_N'(t)$.

More precisely, we show the following result:

\begin{proposition}\label{prop:chder}
	Let $a$ and $b$ be as in~\eqref{eq:ab}. Then,
	\begin{equation}\label{eq:chder}
	\sigma'_N\left(\frac{1}{\tau-1}\right)=-2\left(\frac{A+\frac{3-\tau}{\tau-2}C}{A+\frac{4-\tau}{\tau-2}C}-\frac{D}{E+D}\right),
	\end{equation}
	where
	\begin{align}
	A&=\frac{1}{b^2}\Big(\frac{-\ln(b^2)}{(\tau-1)(\tau-2)}-\frac{1-b^{2(1-\tau)}}{(\tau-1)^2}+\quad \frac{b^{2(\tau-2)}-1}{(\tau-2)^2}\Big),\label{eq:ABCDA}\\
	C&=\left(\frac{b^{\tau-3}-a^{3-\tau}}{3-\tau}\right)^2,\label{eq:ABCDC}\\
	D&=\frac{1}{b}\frac{b^{\tau-1}-b^{1-\tau}}{\tau-1},\label{eq:ABCDD}\\
	E&=\frac{a^{2-\tau}-b^{\tau-2}}{\tau-2}.\label{eq:ABCDE}
	\end{align}
	Furthermore,
	\begin{equation}\label{eq:derlimit}
	\sigma_N'(\tfrac{1}{\tau-1})>\lim_{M\to\infty}\sigma_M'\left(\tfrac{1}{\tau-1}\right)=-2(3-\tau)
	\end{equation}
	for all $N$.
\end{proposition}

The limiting value in~\eqref{eq:derlimit} is consistent with the limiting value of $\sigma_N(t)$ that has been found in~\eqref{eq:sigmalim}. We assess this convergence result with plots. While these indicate that the limits are only reached for very large $N$, especially when $\tau$ is close to 2, it can also be seen that the limiting shape of $\sigma_N(t)$ already shows up for considerably smaller $N$.

To start the proof of Proposition~\ref{prop:chder}, note that in the $a,b$ notation of~\eqref{eq:ab},
\begin{equation}\label{eq:chquot}
c(h)=\frac{K(h)}{J(h)}, \quad 0\leq h\leq h_c,
\end{equation}
where
\begin{align}
K(h)&=\int_{a}^{b}\int_{a}^{b}(xy)^{2-\tau}f(ahx)f(ahy)f(xy)\dd x \dd y,\\
J(h)&=\Big(\int_{a}^{b}x^{1-\tau}f(ahx)\dd x\Big)^2\label{eq:Jh},
\end{align}
with $f(u)=\min(1,u^{-1})$. Note that $r(u)=uf(u)$, see~\eqref{eq:r}. We compute
\begin{equation}\label{eq:chdercomp}
\begin{aligned}[b]
\sigma_N'(t)&=\frac{\dd}{\dd t}\left(\frac{\ln\left(c((N\mean{h})^t)/c(h_{\text{ref}})\right)}{\ln(N\mean{h})}\right) \\
& = (N\mean{h})^t\ln(N\mean{h})\frac{c'((N\mean{h})^t)}{c((N\mean{h})^t)\ln(N\mean{h})}\\
& = h\frac{c'(h)}{c(h)},\quad h=(N\mean{h})^t,
\end{aligned}
\end{equation}
where the prime on $c$ indicates differentiation with respect to $h$. With~\eqref{eq:chquot} we get
\begin{equation}
\frac{c'(h)}{c(h)}=\frac{K'(h)}{K(h)}-\frac{J'(h)}{J(h)},
\end{equation}
and we have to evaluate $K(h), K'(h), J(h)$ and $J'(h)$ at
\begin{equation}\label{eq:hc}
h=h_c=b/a.
\end{equation}

\begin{lemma}\label{lem:ABCD}
	\begin{align}
	K(h_c)&=A+\frac{4-\tau}{2-\tau}C, \ K'(h_c)=\tfrac{-2a}{b}\left(A+\frac{3-\tau}{\tau-2}C\right),\label{eq:Khc}\\
	J(h_c)&=(D+E)^2, \ J'(h_c)=-\tfrac{2a}{b}(D+E)D\label{eq:Jhc},
	\end{align}
	with $A,C,D,E$ as in~\eqref{eq:ABCDA}--\eqref{eq:ABCDE}.
\end{lemma}

From Lemma~\ref{lem:ABCD},~\eqref{eq:chdercomp} and~\eqref{eq:hc} we get~\eqref{eq:chder} in Proposition~\ref{prop:chder}.
\\\\

\textit{Proof of Lemma~\ref{lem:ABCD}.}
	Since $h_c=b/a$,
	\begin{equation}
	K(h_c)=\int_{a}^{b}\int_{a}^{b}(xy)^{2-\tau}f(bx)f(by)f(xy)\dd x \dd y.
	\end{equation}
	With $f(u)=\min(1,u^{-1})$ we split up the integration range $[a,b]\times[a,b]$ into the four regions $[a,1/b]\times[a,1/b]$, $[1/b,b]\times[1/b,b],[1/b,b]\times[a,1/b]$ and $[a,1/b]\times[1/b,b]$, where we observe that $a\leq 1/b\leq 1\leq b$. We first get
	\begin{equation}\label{eq:a1b}
	\begin{aligned}[b]
	&\int_{a}^{1/b}\int_{a}^{1/b}(xy)^{2-\tau}f(bx)f(by)f(xy)\dd x \dd y\\
	& =\int_{a}^{1/b}\int_{a}^{1/b}(xy)^{2-\tau}\cdot 1\cdot 1\cdot 1\dd x \dd y\\
	& =\left(\frac{b^{\tau-3}-a^{3-\tau}}{3-\tau}\right)^2=C.
	\end{aligned}
	\end{equation}
	Next,
	\begin{equation}\label{eq:1bb}
	\begin{aligned}[b]
	&\int_{1/b}^{b}\int_{1/b}^{b}(xy)^{2-\tau}f(bx)f(by)f(xy)\dd x \dd y\\
	& =\int_{1/b}^{b}\int_{1/b}^{b}(xy)^{2-\tau}\frac{1}{bx}\frac{1}{by}f(xy)\dd x \dd y\\
	& =\frac{1}{b^2}\int_{1/b}^{b}\int_{1/b}^{b}(xy)^{1-\tau}f(xy)\dd x \dd y.
	\end{aligned}
	\end{equation}
	The remaining double integral with $\tau+1$ instead of $\tau$ has been evaluated in~\cite[Appendix C, (C3)]{hofstad2017b} as
	\begin{equation}
	-\frac{\ln(b^2)}{(\tau-1)(\tau-2)}-\frac{1-b^{2(1-\tau)}}{(\tau-1)^2}+\frac{b^{2(\tau-2)-1}}{(\tau-2)^2}=b^2A.
	\end{equation}
	Finally, the two double integrals over $[1/b,b]\times[a,1/b]$ and $[a,1/b]\times[1/b,b]$ are by symmetry both equal to
	\begin{equation}\label{eq:1ba}
	\begin{aligned}[b]
	& \int_{1/b}^{b}\int_{a}^{1/b}(xy)^{2-\tau}f(bx)f(by)f(xy)\dd x \dd y \\
	& =\int_{1/b}^{b}\int_{a}^{1/b}(xy)^{2-\tau}\frac{1}{bx}\cdot 1\cdot 1\dd x \dd y\\
	& =\frac{1}{b}\frac{b^{\tau-2}-b^{2-\tau}}{\tau-2}\frac{b^{\tau-3}-a^{3-\tau}}{3-\tau} =\frac{(b^{\tau-3}-a^{3-\tau})^2}{(\tau-2)(3-\tau)}\\
	& =\frac{3-\tau}{\tau-2}C.
	\end{aligned}
	\end{equation}
	Here we have used that, see~\eqref{eq:ab},
	\begin{equation}\label{eq:abrel}
	b^{1-\tau}=a^{3-\tau}.
	\end{equation}
	Now the expression in~\eqref{eq:Khc} for $K(h_c)$ follows.
	
	To evaluate $K'(h_c)$, we observe by symmetry that
	\begin{equation}
	K'(h)=2\int_{a}^{b}\int_{a}^{b}(xy)^{2-\tau}axf'(ahx)f(ahy)f(xy)\dd x  \dd y.
	\end{equation}
	At $h=h_c$, we have $ah=b$, and so
	\begin{equation}
	K'(h_c)=2\frac{a}{b}\int_{a}^{b}\int_{a}^{b}(xy)^{2-\tau}bxf'(bx)f(by)f(xy)\dd x  \dd y.
	\end{equation}
	Now $uf'(u)=0$ for $0\leq u\leq1$ and $uf'(u)=-f(u)$ for $u\geq 1$. Hence, splitting up the integration range into the four regions as earlier, we see that those over $[a,1/b]\times[a,1/b]$ and $[a,1/b]\times[1/b,b]$ vanish while those over $[1/b,b]\times[1/b,b]$ and $[1/b,b]\times[a,1/b]$ give rise to the same double integrals as in~\eqref{eq:1bb} and~\eqref{eq:1ba} respectively. This yields the expression in~\eqref{eq:Khc} for $K'(h_c)$.
	
	The evaluation of $J(h_c)$ and $J'(h_c)$ is straightforward from~\eqref{eq:Jh} with $ah=b$ and a splitting of the integration range $[a,b]$ into $[a,1/b]$ and $[1/b,b]$. This yields~\eqref{eq:Jhc}, and the proof of Lemma~\ref{lem:ABCD} is complete.
	
	We now turn to the limiting behavior of $\sigma_N'(\tfrac{1}{\tau-1})$ as $N\to\infty$. For this we write
	\begin{equation}
	0<\frac{D}{D+E}=\frac{1-b^{2(1-\tau)}}{\frac{\tau-1}{\tau-2}(ab)^{2-\tau}-\frac{1}{\tau-2}-\frac{1}{\tau-1}b^{2(1-\tau)}},
	\end{equation}
	in which
	\begin{align}
	b^{2(1-\tau)}&=(N\mean{h})^{\tau-3}\to 0,\\
	(ab)^{2-\tau}&=(N\mean{h})^{\frac{(\tau-2)^2}{\tau-1}}\to\infty\label{eq:abcon},
	\end{align}
	as $N\to\infty$. Hence, $D/(D+E)\to 0$ as $N\to\infty$. Furthermore, we write
	\begin{equation}\label{eq:ceq}
	C=\frac{b^{2(\tau-3)}}{(\tau-3)^2}\left(1-(ab)^{3-\tau}\right)^2,
	\end{equation}
	and
	\begin{equation}\label{eq:aeq}
	A=\frac{b^{2(\tau-3)}}{(\tau-2)^2}(1-F),
	\end{equation}
	where
	\begin{equation}\label{eq:Feq}
	\begin{aligned}[b]
	F&=b^{-2(\tau-2)}\Big[\frac{\tau-2}{\tau-1}\ln(b^2)\\
	&\quad +\left(\frac{\tau-2}{\tau-1}\right)^2(1-b^{2(1-\tau)})+1\Big] \\
	&= \frac{1}{\tau-1}b^{-2(\tau-2)}\ln(b^{2(\tau-2)})\left(1+O\left(\frac{1}{\ln(b)}\right)\right).
	\end{aligned}
	\end{equation}
	Now, using~\eqref{eq:abrel}, we have
	\begin{equation}\label{eq:chspeed}
	(ab)^{3-\tau}=b^{-2(\tau-2)}=(N\mean{h})^{\frac{(\tau-2)(3-\tau)}{\tau-1}}\to 0
	\end{equation}
	as $N\to\infty$. Thus, we get
	\begin{equation}\label{eq:limder}
	\begin{aligned}[b]
	\lim_{N\to\infty}& \frac{A+\frac{3-\tau}{2-\tau}C}{A+\frac{4-\tau}{2-\tau}C}
	=\frac{\frac{1}{(\tau-2)^2}+\frac{3-\tau}{\tau-2}\frac{1}{(3-\tau)^2}}{\frac{1}{(\tau-2)^2}+\frac{4-\tau}{\tau-2}\frac{1}{(3-\tau)^2}}=3-\tau,
	\end{aligned}
	\end{equation}
	and this yields~\eqref{eq:derlimit}.
	
	Note that $D/(D+E)$ approaches 0 much slower than the limit in~\eqref{eq:limder} is reached when $\tau$ is close to 2, compare~\eqref{eq:abcon} and~\eqref{eq:limder}. Thus, we can concentrate on $D/(D+E)$, and the relative deviation of $\sigma_N'(t)$ from $-2(3-\tau)$ is approximately
	\begin{equation}
		\begin{aligned}[b]
		\frac{2D}{D+E}\frac{1}{2(3-\tau)}&\approx \frac{\tau-2}{3-\tau}\frac{1}{(ab)^{2-\tau}-1}\\
		& \approx \frac{\tau-2}{3-\tau} (N\mean{h})^{-\frac{(\tau-2)^2}{\tau-1}}.
		\end{aligned}
	\end{equation}

\begin{figure*}[htb]
	\centering
	\subfloat[]{
		\centering
		\includegraphics*[width=0.4\textwidth]{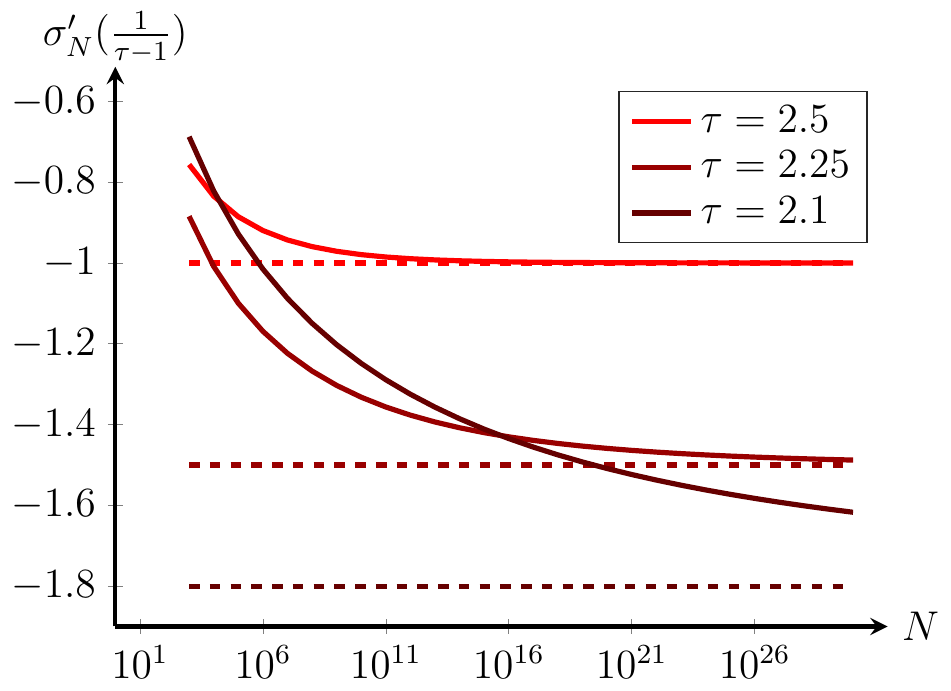}
		\label{fig:derivend}
	}
	\hspace{0.2cm}
	\subfloat[]{
		\centering
		\includegraphics*[width=0.4\textwidth]{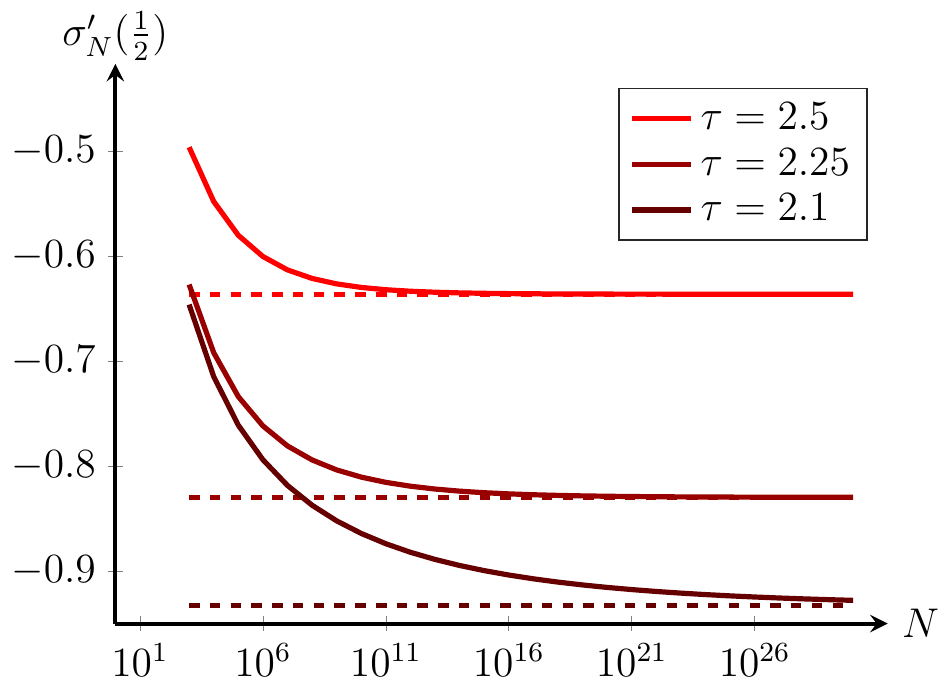}
		\label{fig:derivhalf}
	}
	\caption{$\sigma'_N(t)$ plotted against $N$ for (a) $t=\tfrac{1}{\tau-1}$ and (b) $t=\tfrac{1}{2}$. The dashed line gives the limiting value of $\sigma'_N(t)$ as $N\to\infty$.}
	\label{fig:sigmaprime}
\end{figure*}
	We finally turn to the inequality in~\eqref{eq:derlimit} in Proposition~\ref{prop:chder}. Obviously, we have
	\begin{equation}
	\sigma_N'\left(\tfrac{1}{\tau-1}\right)> -2 \frac{A+\frac{3-\tau}{\tau-2}C}{A+\frac{4-\tau}{\tau-2}C}.
	\end{equation}
	We shall show that
	\begin{equation}\label{eq:ineqder}
	\begin{aligned}[b]
	&\frac{A+\frac{3-\tau}{\tau-2}C}{A+\frac{4-\tau}{\tau-2}C} \leq\frac{A_{\text{as}}+\frac{3-\tau}{\tau-2}C_\text{as}}{A_{\text{as}}+\frac{4-\tau}{\tau-2}C_\text{as}}=3-\tau,
	\end{aligned}
	\end{equation}
	where
	\begin{equation}
	A_{\text{as}}=\frac{b^{2(\tau-3)}}{(\tau-2)^2},\quad C_{\text{as}}=\frac{b^{2(\tau-3)}}{(3-\tau)^2},
	\end{equation}
	the asymptotic form  of $A$ and $C$ as $N\to\infty$ obtained from~\eqref{eq:aeq} and~\eqref{eq:ceq} by deleting $F$ and $(ab)^{3-\tau}$, respectively. The function
	\begin{equation}
	x\in[0,\infty)\mapsto \frac{1+\frac{3-\tau}{\tau-2}x}{1+\frac{4-\tau}{\tau-2}x}
	\end{equation}
	is decreasing in $x\geq 0$, and so it suffices to show that
	\begin{equation}\label{eq:CAas}
	\frac{C_{\text{as}}}{A_{\text{as}}}\leq \frac{C}{A}, \text{ i.e., that }\frac{C_{\text{as}}}{C}\leq\frac{A_{\text{as}}}{A}.
	\end{equation}
	We have from~\eqref{eq:ceq} that
	\begin{equation}
	\frac{C_{\text{as}}}{C}=\frac{1}{\left(1-(ab)^{3-\tau}\right)^2},
	\end{equation}
	and from~\eqref{eq:aeq} and~\eqref{eq:Feq} that
	\begin{equation}
	\begin{aligned}[b]
	\frac{A}{A_\text{as}}&={1-F} =1-b^{-2(\tau-2)}-b^{-2(\tau-2)}\\
	&\times \left[\frac{\tau-2}{\tau-1}\ln(b^2)+\Big(\frac{\tau-2}{\tau-1}\Big)^2(1-b^{2(1-\tau)})\right].
	\end{aligned}
	\end{equation}
	Using that $(ab)^{3-\tau}=b^{-2(\tau-2)}$, see~\eqref{eq:chspeed}, we see that the inequality $C_{\text{as}}/C\leq A_{\text{as}}/A$ in~\eqref{eq:CAas} is equivalent to
	\begin{equation}\label{eq:endeq}
	\begin{aligned}[b]
	&(1-b^{-2(\tau-2)})^2\geq 1-b^{-2(\tau-2)}-b^{-2(\tau-2)}\\
	&\times \left[\frac{\tau-2}{\tau-1}\ln(b^2)+\Big(\frac{\tau-2}{\tau-1}\Big)^2(1-b^{2(1-\tau)})\right].
	\end{aligned}
	\end{equation}
	Using that $(1-u)^2-(1-u)=-u(1-u)$ and dividing through by $u=b^{-2(\tau-2)}$, we see that~\eqref{eq:endeq} is equivalent to
	\begin{equation}\label{eq:ueq}
	\frac{\tau-2}{\tau-1}\ln(b^2)+\Big(\frac{\tau-2}{\tau-1}\Big)^2(1-b^{2(1-\tau)})\geq 1-b^{-2(\tau-2)}.
	\end{equation}
	With $y=\ln(b^2)\geq 0$, we write~\eqref{eq:ueq} as
	\begin{equation}
	K(y):=\Big(\frac{\tau-2}{\tau-1}\Big)^2(1-\me^{(1-\tau)y})+\frac{\tau-2}{\tau-1}y-(1-\me^{(2-\tau)y})\geq 0.
	\end{equation}
	Taylor development  of $K(y)$ at $y=0$ yields
	\begin{equation}
	K(y)=0\cdot y^0+0\cdot y^1+0\cdot y^2+\frac{1}{6}(\tau-2)^2y^3+\dots.
	\end{equation}
	Furthermore,
	\begin{equation}
	K''(y)=(\tau-2)^2\me^{(1-\tau)y} (\me^y-1)> 0,\quad y>0.
	\end{equation}
	Therefore, $K(0)=K'(0)=0$, while $K''(y)>0$ for $y>0$. This gives $K(y)>0$ when $y>0$, as required.

Similar to Proposition~\ref{prop:chder}, we can derive the following result for $\sigma_N'(\tfrac{1}{2})$:
\begin{proposition}
	\begin{equation}
		\sigma'_N(\tfrac12)=-2\Bigg(\frac{G+H}{\Big(1+\big(\frac{\tau-1}{3-\tau}\big)^2\Big)G+2H}-\frac{I}{I+J}\Bigg),
	\end{equation}
	where
	\begin{align}
		G&=\left(\frac{1-b^{1-\tau}}{\tau-1}\right)^2,\\
		I&=\frac{1-b^{1-\tau}}{\tau-1},\\
		J&=\frac{b^{(\tau-2)(\tau-1)/(3-\tau)}-1}{\tau-2},\\
		H&=\frac{1-1/b-b^{1-\tau}(1-b^{2-\tau})}{(\tau-2)(3-\tau)}-\frac{1-b^{1-\tau}}{(\tau-1)(\tau-2)}.
	\end{align}
	Furthermore,
	\begin{equation}
	\sigma_N'(\tfrac 1 2 )> \lim_{M\to\infty}\sigma_M'(\tfrac{1}{2})=-1+\frac{2(\tau-2)}{3-(\tau-2)^2},
	\end{equation}
	for all $N$.
\end{proposition}

Figure~\ref{fig:sigmaprime} shows the values of $\sigma_N'(\tfrac{1}{2})$ and  $\sigma_N'(\tfrac{1}{\tau-1})$ for finite-size networks together with its limiting value.
For example, when $\tau=2.25$, Fig.~\ref{fig:derivend} shows that $N$ needs to be of the order $10^{16}$ for the slope to be `close' to its limiting value -1.5. When for example $N=10^6$ we see that the slope is much smaller: approximately -1.1. This makes statistical estimation of the true underling power-law exponent $\alpha$ extremely challenging, especially for the relevant regime $\tau$ close to $2$, because enormous amounts of data should be available to get sufficient statistical accuracy. Most data sets, even the largest available networks used in this paper, are simply not large enough to have sufficiently many samples from the large-degree region to get a statistically accurate estimate of the power-law part. This also explains why based on smaller data sets it is common to assume that $\alpha$ is roughly one \cite{vazquez2002,ravasz2003,serrano2006b,catanzaro2004, leskovec2008,krioukov2012}. Comparing Fig.~\ref{fig:derivend} and Fig.~\ref{fig:derivhalf} shows that the convergence to the limiting value is significantly faster at the point $t=\tfrac{1}{2}$ than at the point $t=\tfrac{1}{\tau-1}$.

\section{From hidden variables to degrees}\label{sec:hk}
\begin{figure}[htb]
	\centering
	\includegraphics[width=0.4\textwidth]{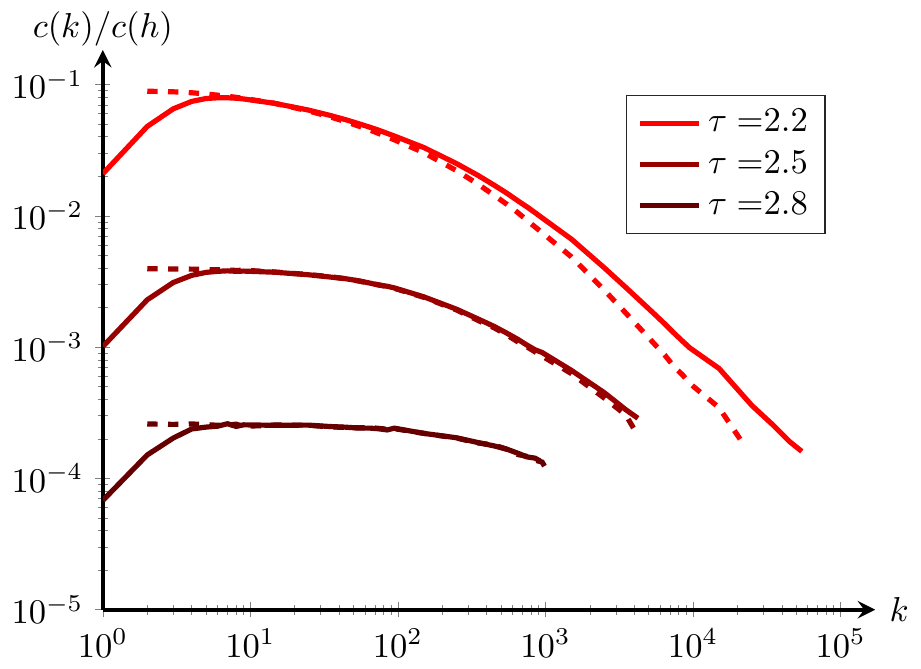}
	\caption{$\bar{c}(k)$ (dashed) and $c(h)$ (line) for $N=10^5$, averaged over $10^4$ realizations.}
	\label{fig:chck}
\end{figure}
In this paper, we focus on computing $c(h)$, the local clustering coefficient of a randomly chosen vertex with hidden variable $h$. However, when studying local clustering in real-world data sets, we can only observe $\bar c(k)$, the local clustering coefficient of a vertex of degree $k$. In this appendix, we show that for the hidden variable model, the difference between these two methods of computing the clustering coefficient is small and asymptotically negligible.
We consider
\begin{equation}\label{eq:cab}
c(h) = \frac{\int_{1}^{h_c}\int_{1}^{h_c}(h'h'')^{2-\tau}p(h,h')p(h, h'')p(h',h'')\dd h'\dd h''}{\left(\int_{1}^{h_c}x^{1-\tau}p(h, h')\dd h'\right)^2}.
\end{equation}
We define $\bar{c}(k)$ as the average clustering coefficient over all vertices of degree $k$. By~\cite{colomer2012}, the probability that a vertex with hidden variable $h$ has degree $k$ equals
\begin{equation}
g(k\mid h) = \frac{\me^{-h}h^k}{k!}.
\end{equation}
Then, by~\cite{colomer2012},
\begin{equation}
\bar{c}(k) =\begin{cases}
\frac{1}{P(k)}\int_{1}^{h_c}\rho(h)c(h)g(k\mid h)\dd h, & \quad k\geq 2,\\
0, & \quad k<2,
\end{cases}
\end{equation}
where $\bar{c}(k)=0$ for $k<2$ because a vertex with degree less than 2 cannot be part of a triangle.
Here
\begin{equation}
P(k) = \int_{1}^{h_c}g(k\mid h)\rho(h)\dd h
\end{equation}
is the probability that a randomly chosen vertex has degree $k$.

First we consider the case where $h> N^{\frac{\tau-2}{\tau-1}}$.
The Chernoff bound gives for the tails of the Poisson distribution that
\begin{align}
\Prob{\text{Poi}(\lambda)>x} & \leq \me^{-\lambda}\left(\frac{\me \lambda}{x}\right)^x,\quad x>\lambda,  \\
\Prob{\text{Poi}(\lambda)<x} & \leq \me^{-\lambda}\left(\frac{\me \lambda}{x}\right)^x ,\quad x<\lambda.
\end{align}
Let $k(h)$ be the degree of a node with hidden variable $h$. Then, for any $M>1$
\begin{equation}\label{eq:klarg}
\sum_{k=Mh}^{\infty}g(k\mid h)\leq \left(\frac{\me^{M-1}}{M^M}\right)^h,
\end{equation}
and for any $\delta\in(0,1)$,
\begin{equation}\label{eq:ksmall}
\sum_{k=1}^{\delta h}g(k\mid h)\leq \left(\frac{\me^{\delta-1}}{\delta^\delta}\right)^h.
\end{equation}
Because $\me^{x-1}/x^x<1$ for $x\neq 1$,~\eqref{eq:klarg} and~\eqref{eq:ksmall} tend to zero as $h\to\infty$. Therefore, for $h$ large,
\begin{equation}
k(h)= h(1+o(1))
\end{equation}
with high probability. Therefore, when $k$ is large,
\begin{equation}
\bar{c}(k)\approx  c(k).
\end{equation}
Thus, $c(h)$ is very similar to $\bar{c}(k)$.

On the other hand, for $h\ll \hst/h_c$,
\begin{equation}
\sum_{\hst/h_c}^\infty g(k\mid h)\leq \me^{-h}\left(\frac{\me h}{\hst/h_c}\right)^{\hst/h_c},
\end{equation}
which is small by the assumption on $h$. Thus,
\begin{equation}
P(k)\approx \int_{1}^{\hst/h_c}g(k\mid h)\rho(h)\dd h.
\end{equation}
Furthermore, $c(h)=c(0)$ in this regime of $h$. This results in
\begin{equation}
\bar c (k)\approx \frac{c(0)\int_{1}^{\hst/h_c}\rho(h)g(k\mid h)\dd h}{\int_{1}^{\hst/h_c}\rho(h)g(k\mid h)\dd h}=c(0).
\end{equation}
Therefore, $\bar{c}(h)\approx c(h)$ also when $h$ is small.

Figure~\ref{fig:chck} shows that indeed the difference between $\bar{c}(k)$ and $c(k)$ is small. When $\tau$ approaches 2, the difference becomes larger. We see that for small values of $k$, $\bar{c}(k)$ and $c(k)$ are not very close. This is due to the fact that~\eqref{eq:cab} does not take into account that a vertex with hidden variable $h$ may have less than 2 neighbors, so that its local clustering is zero. In~\cite{hofstad2017b} we show how to adjust~\eqref{eq:chab} to account for this.

\section{Degree distributions}\label{sec:degree}
\begin{figure*}[htb]
	\centering
	\subfloat[]{
		\centering
		\includegraphics*[width=0.4\textwidth]{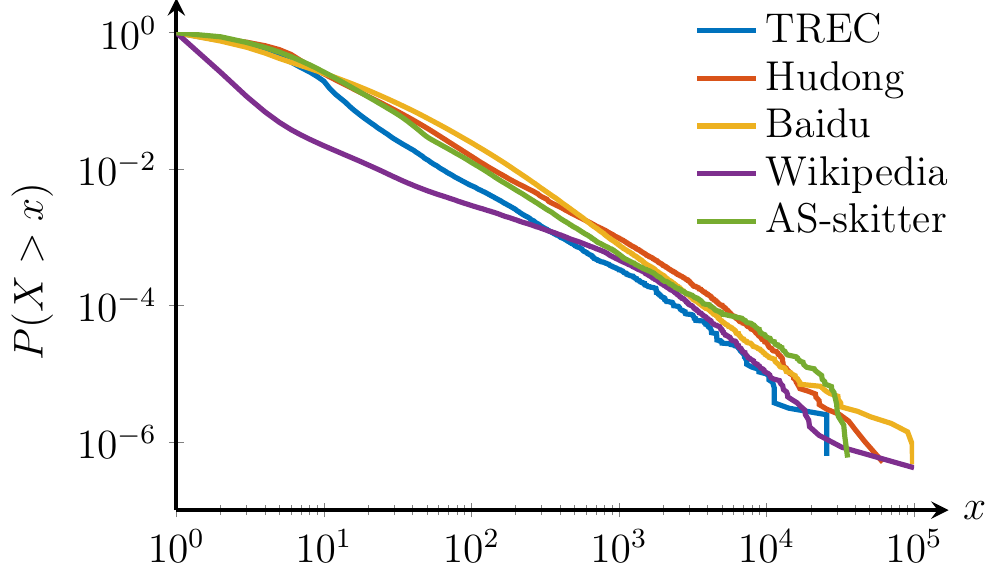}
		\label{fig:deg1}
	}
	\hspace{0.2cm}
	\subfloat[]{
		\centering
		\includegraphics*[width=0.4\textwidth]{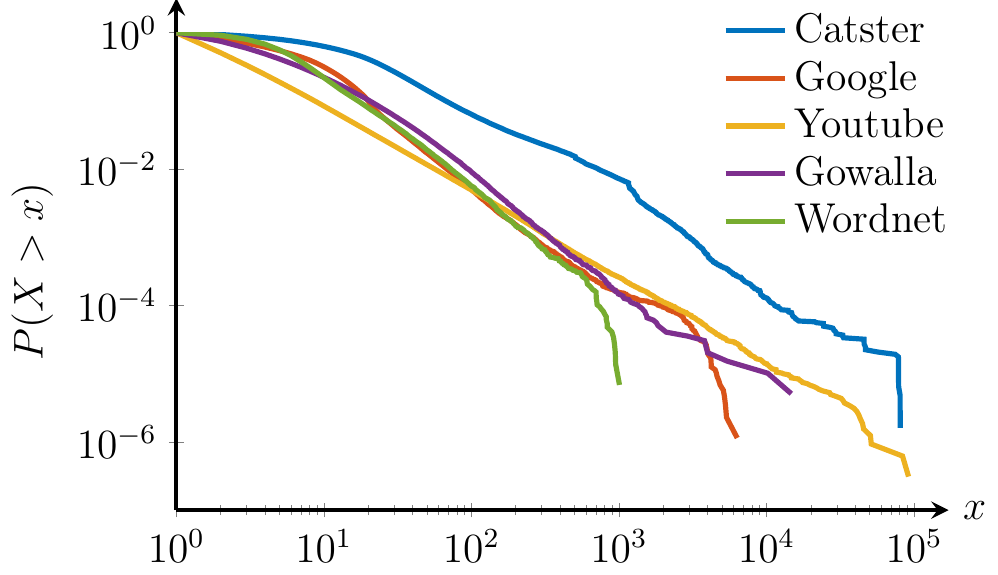}
		\label{fig:deg2}
	}
	\caption{The probability that the degree of a vertex exceeds $x$ in a) the largest 5 networks of Table~\ref{tab:data} b) the smallest 5 networks in Table~\ref{tab:data}}
	\label{fig:degree}
\end{figure*}

Figure~\ref{fig:degree} shows the degree distributions of all ten networks of Table~\ref{tab:data}.
\cleardoublepage

\end{document}